\documentclass[twocolumn,times]{aastex631}

\newcommand{\teff}{${T}_{\mathrm{eff}}$}
\newcommand{\logg}{$\log{g}$}
\newcommand{\msun}{$M_{\odot}$}

\newcommand{\lsun}{$L_{\odot}$}
\newcommand{\kms}{km\,s$^{-1}$}

\newcommand{\muhz}{${\rm \mu}$Hz}
\newcommand{\bvf}{Brunt-V\"{a}is\"{a}l\"{a}}

\accepted{October 4, 2021}

\submitjournal{ApJ}

\shorttitle{Pulsations in the ELM WD GD 278}
\shortauthors{Lopez et al.}

\begin{document}

\title{Discovery, {\em TESS} Characterization, and Modeling of Pulsations in the Extremely Low Mass White Dwarf GD 278}

\correspondingauthor{Isaac D. Lopez}
\email{ilopez1@highpoint.edu}

\author[0000-0002-0009-409X]{Isaac D. Lopez}
\affiliation{High Point University, Department of Physics, One University Parkway, High Point, NC 27268, USA}
\affiliation{Department of Astronomy \& Institute for Astrophysical Research, Boston University, 725 Commonwealth Ave., Boston, MA 02215, USA}

\author[0000-0001-5941-2286]{J. J. Hermes}
\affiliation{Department of Astronomy \& Institute for Astrophysical Research, Boston University, 725 Commonwealth Ave., Boston, MA 02215, USA}

\author[0000-0002-1345-8075]{Leila M. Calcaferro}
\affiliation{Grupo de Evoluci\'{o}n Estelar y Pulsaciones, Facultad de Ciencias Astron\'{o}micas y Geof\'{i}sicas, Universidad Nacional de La Plata, Paseo del Bosque s/n, 1900, La Plata, Argentina}
\affiliation{Instituto de Astrof\'{i}sica La Plata, CONICET-UNLP, Paseo del Bosque s/n, 1900, La Plata, Argentina}

\author[0000-0002-0656-032X]{Keaton J.\ Bell}
\altaffiliation{NSF Astronomy and Astrophysics Postdoctoral Fellow}
\affil{DIRAC Institute, Department of Astronomy, University of Washington, Seattle, WA 98195, USA}

\author[0000-0003-1895-2934]{Adam Samuels}
\affiliation{Department of Astronomy \& Institute for Astrophysical Research, Boston University, 725 Commonwealth Ave., Boston, MA 02215, USA}

\author[0000-0002-0853-3464]{Zachary P. Vanderbosch}
\affiliation{Department of Astronomy, 
University of Texas at Austin, Austin, TX-78712, USA}

\author[0000-0002-0006-9900]{Alejandro H. C{\'o}rsico}
\affiliation{Grupo de Evoluci\'{o}n Estelar y Pulsaciones, Facultad de Ciencias Astron\'{o}micas y Geof\'{i}sicas, Universidad Nacional de La Plata, Paseo del Bosque s/n, 1900, La Plata, Argentina}
\affiliation{Instituto de Astrof\'{i}sica La Plata, CONICET-UNLP, Paseo del Bosque s/n, 1900, La Plata, Argentina}

\author[0000-0002-8811-8171]{Alina G. Istrate}
\affiliation{Department of Astrophysics/IMAPP, Radboud University, PO Box 9010, 6500, GL Nijmegen, The Netherlands}

\begin{abstract}

We report the discovery of pulsations in the extremely low mass (ELM), likely helium-core white dwarf GD\,278 via ground- and space-based photometry. GD\,278 was observed by the \emph{Transiting Exoplanet Survey Satellite} (\emph{TESS}) in Sector 18 at a 2-min cadence for roughly 24\,d. The \emph{TESS} data reveal at least 19 significant periodicities between $2447-6729$\,s, one of which is the longest pulsation period ever detected in a white dwarf. Previous spectroscopy found that this white dwarf is in a 4.61 hr orbit with an unseen $>$0.4\,\msun\ companion and has ${T}_{\mathrm{eff}} = 9230 \pm 100$\,K and $\log{g} = 6.627 \pm 0.056$, which corresponds to a mass of $0.191\pm0.013$\,$M_{\odot}$. Patterns in the {\em TESS} pulsation frequencies from rotational splittings appear to reveal a stellar rotation period of roughly 10\,hr, making GD 278 the first ELM white dwarf with a measured rotation rate. The patterns inform our mode identification for asteroseismic fits, which unfortunately do not reveal a global best-fit solution. Asteroseismology reveals two main solutions roughly consistent with the spectroscopic parameters of this ELM white dwarf, but with vastly different hydrogen-layer masses; future seismic fits could be further improved by using the stellar parallax. GD\,278 is now the tenth known pulsating ELM white dwarf; it is only the fifth known to be in a short-period binary, but is the first with extended, space-based photometry.

\end{abstract}


\section{Introduction} \label{sec:intro}

White dwarf stars represent the final evolutionary state of main-sequence stars with masses $\lesssim$ 8-10\,\msun, and are the end stages for $\sim$98\% of all stars in the Milky Way. The observed mass distribution of spectroscopically confirmed hydrogen-atmosphere (DA) white dwarfs from the Sloan Digital Sky Survey (SDSS) peaks around 0.6\,\msun, with smaller distributions within both tail ends, the smaller of which peaks around 0.4\,\msun\ (e.g., \citealt{2007MNRAS.375.1315K,2011ApJ...730..128T}). The lower mass limit for isolated white dwarfs is $\sim$0.4\,\msun \citep{2007ApJ...671..761K}, constrained by the age of the Milky Way. Therefore, the progenitors of extremely low mass (ELM; $\lesssim$ 0.3\,\msun) white dwarfs must have been giant stars that experienced enhanced mass loss due to a close companion, the first episode of which occurred on the first ascent of the red-giant branch \citep{1995MNRAS.275..828M, 1998A&A...339..123D}. This process can occur in a close binary, where the progenitor star transfers mass either via Roche lobe overflow \citep{1986ApJ...311..742I} or a common-envelope event \citep{1976IAUS...73...75P, 1993PASP..105.1373I}.

DAV stars (also called ZZ Ceti stars) are variable white dwarfs that undergo pulsations in a narrow temperature range where a convection zone forms, from roughly 12,500\,K to 10,000\,K in canonical-mass (0.6\,\msun) white dwarfs. These global, non-radial $g$-mode pulsations are driven by partially ionized regions that form within their atmospheres. These pulsations manifest as optical variations of up to 30\% and offer the unique opportunity to probe their interiors via asteroseismology \citep{2008ARA&A..46..157W, 2008PASP..120.1043F, 2010A&ARv..18..471A, 2019A&ARv..27....7C}.

Pulsations have previously been observed in nine ELM white dwarfs \citep{2012ApJ...750L..28H, 2013ApJ...765..102H, 2013MNRAS.436.3573H, 2015ASPC..493..217B,2017ApJ...835..180B, 2018A&A...617A...6B, 2018MNRAS.478..867P, 2021ApJ...912..125G}, including one with a millisecond pulsar companion \citep{2015MNRAS.446L..26K, 2018MNRAS.479.1267K}. These discoveries effectively extend the DAV instability strip to lower surface gravities and cooler temperatures (6.0 $\lesssim \log g \lesssim$ 6.8 and 7800 K $\lesssim T_{\rm eff} \lesssim$ 10,000 K). This low-mass extension likely also contains contaminants, especially metal-poor A stars \citep{2018MNRAS.475.2480P, 2018MNRAS.478..867P}, in contrast to the space occupied purely by classical DAVs at higher masses (e.g., \citealt{2017ApJ...835..180B}). 

Most of the known radial-velocity-confirmed ELM white dwarfs were discovered by the ELM Survey \citep{2010ApJ...723.1072B, 2012ApJ...744..142B,2013ApJ...769...66B, 2016ApJ...818..155B, 2020ApJ...889...49B, 2011ApJ...727....3K,  2012ApJ...751..141K,  2015ApJ...812..167G}, a targeted spectroscopic search for ELM white dwarfs, which has discovered 98 double-white-dwarf binaries within the SDSS footprint. An additional eight ELM white dwarfs were reported by \cite{2020ApJ...894...53K}, as part of an extension of the ELM Survey into the southern hemisphere. This southern extension relies on photometry from the VST ATLAS and SkyMapper surveys, and also employs a \emph{Gaia} astrometry-based selection, which chooses candidates from a unique region of parallax-magnitude space occupied by known ELM white dwarfs.   

The population of ELM white dwarfs represent a collection of objects that contribute to a steady foreground of gravitational waves. The sensitivity of the future \emph{Laser Interferometer Space Antenna} (\emph{LISA}) observatory is expected to be set by the noise floor from unresolved double-white-dwarf binaries \citep{2017arXiv170200786A}. Identifying and characterizing compact binary systems in the Milky Way will help characterize the noise floor that may otherwise impede on \emph{LISA}'s ability to detect gravitational waves.

The field of white dwarf astronomy is currently undergoing a transformation thanks to space-based missions such as \emph{Gaia} \citep{2016A&A...595A...1G} and the \emph{Transiting Exoplanet Survey Satellite} (\emph{TESS}; \citealt{2015JATIS...1a4003R}). With the release of \emph{Gaia} DR2, the number of white dwarfs we now know about has exploded from $\sim$33,000 to $\sim$260,000 \citep{2019MNRAS.482.4570G}, including a large catalog of candidate ELM white dwarfs \citep{2019MNRAS.488.2892P}.

\emph{TESS} has completed its initial 2-year mission to observe $\sim$85\% of the celestial sphere in search of small planets orbiting nearby stars. Observations were split over twenty-six 24$\times$96 square degree sectors, each of which were observed for roughly 27 days. \emph{TESS} delivered full-frame images at a 30-min cadence, as well as 2-min cadence data for select objects, as part of their Guest Investigator and Director's Discretionary Targets programs. The 2-min cadence observations give the opportunity for continuous monitoring of bright white dwarfs ($V \lesssim$ 16.5\,mag), which both provide much needed follow-up observations and, due to the near 27-day continuous observations, give excellent opportunities to unambiguously determine periods without aliasing and with precision sufficiently high for reliable asteroseismology. 

In this paper, we present the discovery of pulsations in the ELM white dwarf GD\,278 via high-speed optical photometry. We also present frequency analysis of the \emph{TESS} observations, including asteroseismic modeling of the frequency solution we obtain. The rest of this paper is organized as follows: Section~\ref{sec:observations} describes how GD\,278 was selected for follow up and describes the observations. Section~\ref{sec:analysis} presents the frequency analysis of the \emph{TESS} data. Asteroseismic modeling of GD\,278 is presented in Section~\ref{sec:modeling}. A discussion and conclusions from this discovery are given in Section~\ref{sec:discussion}.

\section{Observations} \label{sec:observations}

GD\,278 (WDJ013058.07+532139.71, $G$\,=\,14.9\,mag) was selected for follow up observations from the \cite{2019MNRAS.482.4570G} \emph{Gaia} DR2 catalog of white dwarfs based on its location on the Hertzsprung-Russell (HR) diagram relative to other known pulsating ELM white dwarfs and on its intrinsic variability, which we estimated from its empirically determined $G$-band flux uncertainty \citep{2021ApJ...912..125G}. Objects that are intrinsically variable (e.g. pulsators, cataclysmic variables, eclipsing binaries) may have anomalously large flux uncertainties at a given magnitude. With this technique we estimate GD\,278 to be among the top 1\% most variable white dwarfs within 200\,pc, prompting us to obtain high-speed photometry. Shortly after our increased interest in GD\,278, it was published as an ELM white dwarf in a 4.61-hr single-lined spectroscopic binary by the ELM Survey \citep{2020ApJ...889...49B}.

\begin{figure*}[ht]
\hspace*{-1cm}
\epsscale{1.1}
\plotone{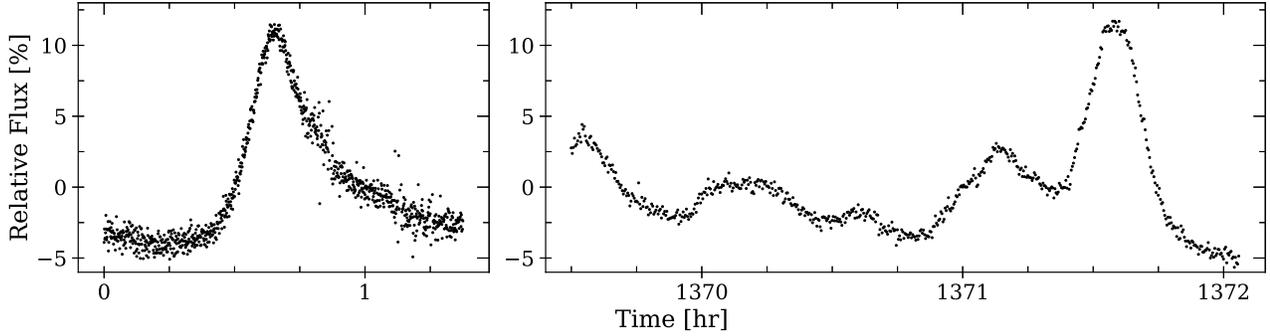}
\caption{High-speed photometry taken with the 2.1 m Otto Struve telescope at McDonald Observatory, identifying pulsations on 2019 August 2 (left) and confirming them on 2019 September 29 (right) in the extremely low-mass white dwarf GD 278. \label{fig:mcd}}
\end{figure*}

\begin{figure*}[ht]
\hspace*{-1cm}
\epsscale{1.1}
\plotone{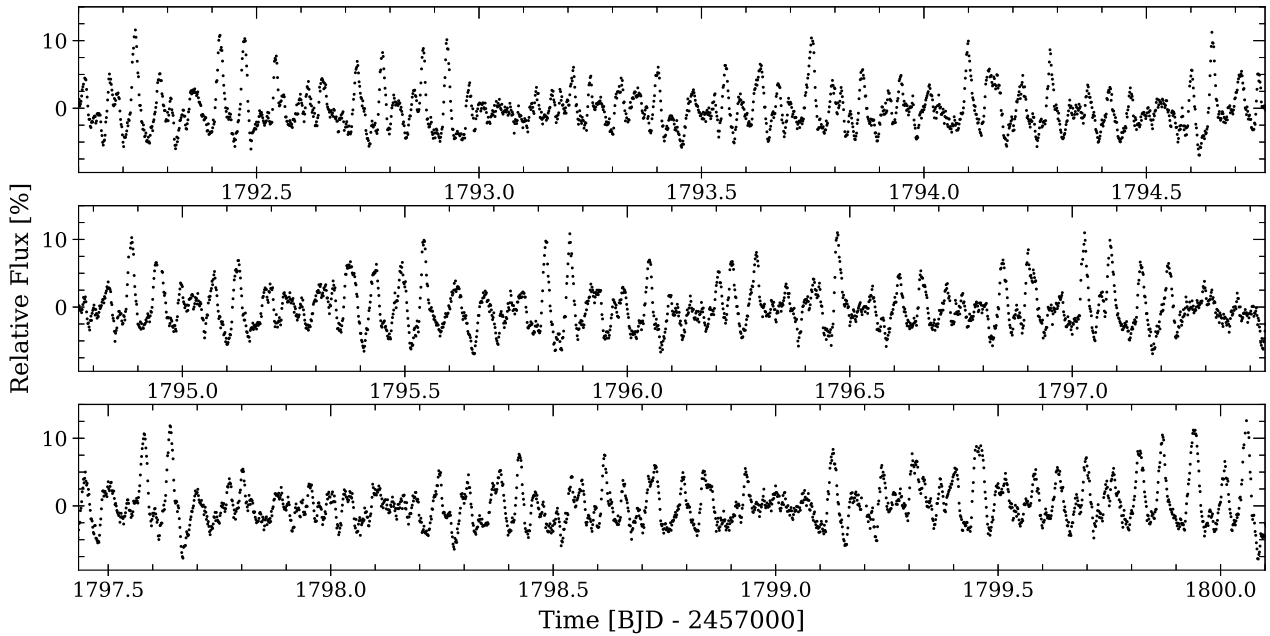}
\caption{An 8-day sample \emph{TESS} photometry of the new pulsating extremely low mass white dwarf GD\,278, which was observed in Sector 18. The 2-minute-cadence data have been smoothed with a 7-point moving average.}
\label{fig:tess}
\end{figure*}

\subsection{McDonald Observations}
Discovery high-speed photometry was obtained with the 2.1 m Otto Struve telescope at McDonald Observatory near Fort Davis, Texas. Observations were obtained over three nights in 2019, all using the frame-transfer ProEM instrument and a red-cutoff {\em BG40} filter to reduce sky noise. Exposure times were 5\,s on 2019 August 2 and 15\,s on 2019 September 29 and 2019 November 6. Skies were clear with 1--2\arcsec\ seeing, with the exception of the 2019 November data, which were affected by poor weather.

Bias, dark, and dome-flat frames were also obtained at the beginning of each night. We reduced the science frames using standard IRAF packages and procedures \citep{1986SPIE..627..733T}. Flux values of the target star were divided by a bright comparison star in order to perform differential photometry. Data from the McDonald observations are shown in Figure~\ref{fig:mcd}. 

Highly discrepant points due to bad weather have been clipped by hand, and the times have been barycentric corrected using the software routine \verb|barycorrpy| \citep{2018RNAAS...2....4K}.

\subsection{\emph{TESS} Observations} \label{subsec:tess}
Based in part on our detection of variability from McDonald Observatory, we submitted a {\em TESS} Director's Discretionary Time (DDT) proposal (PI: Lopez) to secure 2-min photometry of candidate variable ELM white dwarfs. This included GD 278 (TIC 308292831, $T$=14.8\,mag), which was observed by \emph{TESS} in Sector 18 (2019 November 2--27) by Camera 2, temporarily overlapping with our data obtained at McDonald Observatory on 2019 November 6. \emph{TESS} delivered a $11 \times 11$ target pixel file for GD\,278 at a 2-min cadence, with the first mid-exposure time ($t_0$) starting at BJD 2458790.66104. Data was accessed and downloaded using the Python package \verb|lightkurve| \citep{2018ascl.soft12013L}. We tested different apertures for light curve extraction, but found no improvement over the 4-pixel aperture determined by the Science Processing Operations Center (SPOC) pipeline, which we adopt for our analysis.

An 8-day sample of the GD 278 \emph{TESS} photometry is presented in Figure~\ref{fig:tess}. For visual clarity, the data has been smoothed with a 7-point moving average. Crowding is a factor for the large \emph{TESS} plate scale (the SPOC pipeline estimates that 42\% of the flux in the chosen aperture is coming from GD\,278), but this amplitude correction from flux dilution has been applied to our final adopted PDCSAP light curve.

\section{Analysis} \label{sec:analysis}

\subsection{Frequency Solution of the \emph{TESS} Observations}
GD 278 exhibits an irregular pulsation pattern with up to 25$\%$ peak-to-peak brightness variations in both the McDonald and the \emph{TESS} light curves. In order to determine the frequency components contributing to the observed light curve, we employed the Python package \verb|Pyriod|\footnote{\url{http://www.github.com/keatonb/Pyriod}}, a period detection and fitting software that utilizes a Lomb-Scargle periodogram \citep{2020AAS...23510606B}. In Table~\ref{tab:freq_sol}, we list 19 frequencies ($f_a, f_b, ..., f_{s}$) that we detect with significance in the \emph{TESS} data, which we interpret as independent pulsation modes occurring within the white dwarf GD\,278. Properties of these modes that we include in Table~\ref{tab:freq_sol} are their periods, frequencies, amplitudes, phases (relative to the $t_0$ in Section~\ref{subsec:tess}), and associated uncertainties. The dominant frequency $f_a$ is located at 190.36 $\mu$Hz and has an amplitude 6.32 times the significance threshold (described below). We find one combination frequency within this set, occurring at the location $f_a + f_c$, which is likely not a natural oscillation but rather a nonlinear artifact (e.g., \citealt{2001MNRAS.323..248W}).

\begin{deluxetable*}{ccccc}
\tablecaption{The 19 significant pulsation periods/frequencies, one nonlinear combination frequency ($f_a + f_c$), and the orbital frequency ($f_{\mathrm{orb}}$) detected within the GD\,278 \emph{TESS} data, along with their measured amplitudes, phases (relative to BJD 2458790.66104), and associated errors. Each detection has an amplitude above the 0.1\% False Alarm Probability threshold of 0.34\%. \label{tab:freq_sol}}
\tablenum{1}
\tablehead{
\colhead{ID} & 
\colhead{\hspace{.2cm}Period (s)}\hspace{.2cm} &
\colhead{\hspace{.2cm}Frequency ($\mu$Hz)}\hspace{.2cm} &
\colhead{\hspace{.2cm}Amplitude (\%)}\hspace{.2cm} &
\colhead{Phase} }
\tablecolumns{5}
\startdata
$f_a$	&	5253.23  $\pm$ 0.19 	&	190.359	$\pm$	0.007	&	2.15	$\pm$	0.06	&	0.503	$\pm$	0.004	\\
$f_b$	&	3365.56  $\pm$ 0.14	&	297.127	$\pm$	0.012	&	1.25	$\pm$	0.06	&	0.836	$\pm$	0.007	\\
$f_c$	&	4028.83  $\pm$ 0.21 	&	248.211	$\pm$	0.013	&	1.19	$\pm$	0.06	&	0.508	$\pm$	0.007	\\
$f_d$	&	5814.8   $\pm$ 0.7 	&	171.975	$\pm$	0.021	&	0.71	$\pm$	0.06	&	0.528	$\pm$	0.013	\\
$f_e$	&	4628.8   $\pm$ 0.5 	&	216.039	$\pm$	0.022	&	0.67	$\pm$	0.06	&	0.798	$\pm$	0.013	\\
$f_f$	&	5550.6   $\pm$ 0.8	&	180.161	$\pm$	0.025	&	0.61	$\pm$	0.06	&	0.681	$\pm$	0.015	\\
$f_g$	&	6295.4   $\pm$ 1.0 	&	158.846	$\pm$	0.026	&	0.58	$\pm$	0.06	&	0.206	$\pm$	0.015	\\
$f_h$	&	4428.5   $\pm$ 0.5 	&	225.812	$\pm$	0.028	&	0.53	$\pm$	0.06	&	0.012	$\pm$	0.017	\\
$f_i$	&	5198.3   $\pm$ 0.8	&	192.370	$\pm$	0.029	&	0.53	$\pm$	0.06	&	0.986	$\pm$	0.017	\\
$f_j$	&	4259.1   $\pm$ 0.5 	&	234.79	$\pm$	0.03	&	0.50	$\pm$	0.06	&	0.470	$\pm$	0.018	\\
$f_k$	&	5896.9   $\pm$ 1.0 	&	169.58	$\pm$	0.03	&	0.48	$\pm$	0.06	&	0.982	$\pm$	0.018	\\
$f_l$	&	5135.1   $\pm$ 0.8	&	194.74	$\pm$	0.03	&	0.47	$\pm$	0.06	&	0.063	$\pm$	0.019	\\
$f_m$	&	4756.5   $\pm$ 0.7	&	210.24	$\pm$	0.03	&	0.47	$\pm$	0.06	&	0.440	$\pm$	0.019	\\
$f_n$	&	2447.86  $\pm$ 0.18	&	408.52	$\pm$	0.03	&	0.46	$\pm$	0.06	&	0.086	$\pm$	0.019	\\
$f_o$	&	6729.0   $\pm$ 1.8 	&	148.61	$\pm$	0.04	&	0.38	$\pm$	0.06	&	0.865	$\pm$	0.024	\\
$f_p$	&	4959.6   $\pm$ 1.0	&	201.63	$\pm$	0.04	&	0.37	$\pm$	0.06	&	0.483	$\pm$	0.024	\\
$f_q$	&	4093.8   $\pm$ 0.7	&	244.27	$\pm$	0.04	&	0.37	$\pm$	0.06	&	0.115	$\pm$	0.024	\\
$f_r$	&	2658.51  $\pm$ 0.28 	&	376.15	$\pm$	0.04	&	0.35	$\pm$	0.06	&	0.028	$\pm$	0.026	\\
$f_s$	&	3591.3  $\pm$ 0.5	&	278.45	$\pm$	0.05	&	0.34	$\pm$	0.06	&	0.664	$\pm$	0.026	\\
$f_a + f_c$	&	2280.15  $\pm$ 0.15	&	438.567	$\pm$	0.028	&	0.54	$\pm$	0.06	&	0.816	$\pm$	0.017	\\
$f_{\mathrm{orb}}$	&	16611    $\pm$ 8	&	60.20	$\pm$	0.03	&	0.46	$\pm$	0.06	&	0.091	$\pm$	0.019	\\
\enddata
\end{deluxetable*}

\begin{figure}[t]
\hspace*{-0.8
cm}
\epsscale{1.2}
\plotone{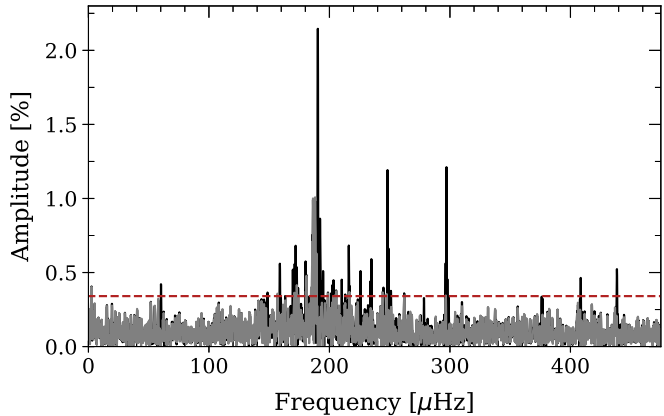}
\caption{Periodogram of the GD 278 \emph{TESS} data in black. In grey is the periodogram of the same data after prewhitening by the frequencies in Table~\ref{tab:freq_sol}. The red dashed line represents the 0.1\% False Alarm Probability threshold, which we estimate to be 0.34\%. Peaks above this line have a $<$0.1\% chance of being caused by random noise, and thus represent intrinsic signals from the star.  \label{fig:ft}}
\end{figure}

The periodogram of the \emph{TESS} data is shown in black in Figure~\ref{fig:ft}, which was used as input to determine the periods in Table~\ref{tab:freq_sol}. In grey we show a periodogram of the \emph{TESS} data after prewhitening by the list of adopted frequencies in Table~\ref{tab:freq_sol}. The red dashed line represents the 0.1\% False Alarm Probability threshold; peaks above this value have a low probability ($<$0.1\%) of resulting from random noise and are likely intrinsic to the star. We bootstrapped the value of this threshold by keeping the time sampling the same but each time randomly selecting fluxes from the set of the flux values. After $10^4$ iterations, the False Alarm Probability threshold was taken to be value of the tenth highest amplitude within the ensemble of periodograms, and has a value of 0.34\%.

In some cases, we chose not to consider certain peaks due their proximity to other peaks with very high amplitudes that we took to be independent modes. For example, we do not consider the two peaks located at 186.4 \muhz\ and 188.1 \muhz, due to their proximity to the peak located at 190.36 \muhz. This mode is one of the lowest-frequency pulsations in GD\,278, which may lose phase coherence in the same way that lower-frequency modes do in normal-mass DA white dwarfs \citep{2017ApJS..232...23H,2020ApJ...890...11M}. Thus, these residual peaks to slightly lower frequencies from $f_{6c}$ may not be independent modes themselves, but rather artifacts from the phase incoherence and amplitude variations of the high-amplitude pulsation.

\subsection{Simultaneous McDonald and \emph{TESS} Observations}
\begin{figure}[t]
\epsscale{1.15}
\plotone{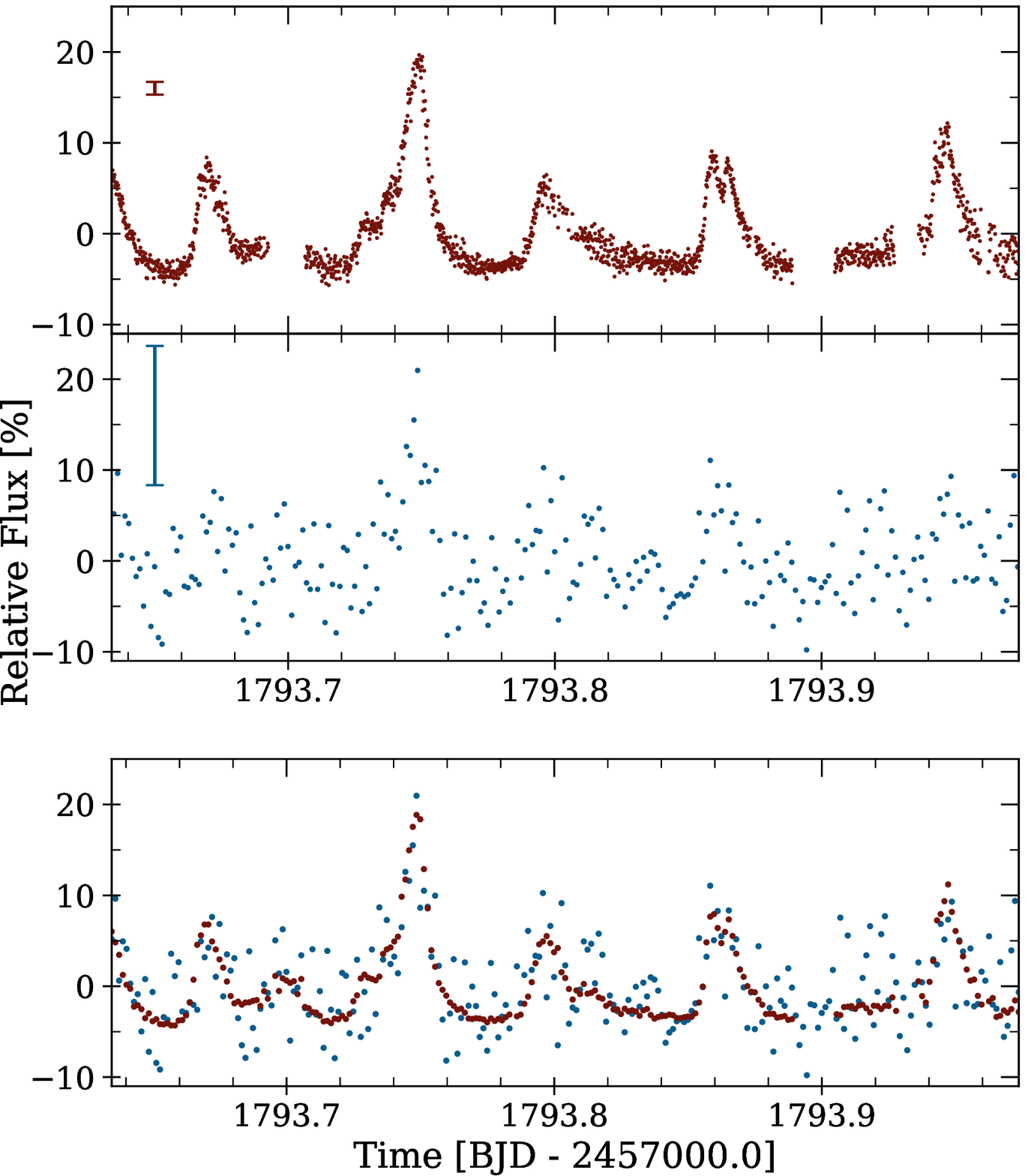}
\caption{{\bf Top:} Simultaneous McDonald (15-s exposures) and \emph{TESS} (120-s exposures) observations of GD 278, captured on 2019 November 8. A representative error bar is shown in the top left corner of each plot. {\bf Bottom:} Both data sets overplotted. The McDonald data (red) have been resampled to the same 2-min cadence as the \emph{TESS} data (blue).   \label{fig:simul}}
\end{figure}

The 2019 November data from McDonald Observatory were obtained simultaneously while GD 278 was being observed by \emph{TESS}. We show the 8.16 hr of overlapping data in Figure~\ref{fig:simul}. Time is given in \emph{TESS} Barycentric Julian Date (TBJD), which is the Julian Date at the Solar System barycenter, offset by 2457000.0. The higher sampling rate (15\,s) and signal-to-noise of the ground-based observations reveal shorter-timescale variations that are missed by the 2-min cadence \emph{TESS} data, especially the double-peaked feature located at roughly 1739.86 TBJD in Figure~\ref{fig:simul}.

For better comparison, we have resampled the McDonald observations at the same 2-min cadence of \emph{TESS}. These data are plotted over each other in bottom panel of Figure~\ref{fig:simul} to highlight that the shortest-period features would be better resolved with future \emph{TESS} observations obtained in the 20-second-cadence mode. It is possible that our frequency solution in Table~\ref{tab:freq_sol} lacks some of the higher-frequency components that contribute to the peaky, nonlinear light curve shape resolved by the McDonald observations. However, such modes are not independent pulsations and should not significantly affect our asteroseismic analysis in Section~\ref{sec:modeling}.

\subsection{Orbital Period Measurement}
In addition to pulsations, we also recover the spectroscopically determined orbital period of 4.6092 $\pm$ 0.0048 hr from \citet{2020ApJ...889...49B} in the \emph{TESS} photometry. We detect a significant signal at 60.21 $\pm$ 0.03 $\mu$Hz, which provides an independent measurement of the orbital period from the \emph{TESS} data of 4.6142 $\pm$ 0.0022 hr, in good agreement with the period determined from spectroscopy and caused by the 209.1 $\pm$ 5.1 \kms\ radial-velocity variability of the ELM white dwarf. This photometric signal, with an amplitude in the \emph{TESS} bandpass of 0.46 $\pm$ 0.06\%, may be caused by Doppler beaming of the white dwarf \citep{2010ApJ...725L.200S, 2014ApJ...792...39H}.

\begin{deluxetable*}{lcrcccc}
\tablecaption{We identify common frequency spacings that may reveal the identity (especially the $\ell$ values) for eight different modes in GD 278; each multiplet is illustrated in Figure~\ref{fig:ell_1} and Figure~\ref{fig:ell_2}. In each subset we have attempted to identify the $m=0$ central component critical for asteroseismic analysis; we use a dagger symbol ($^{\dagger}$) to identify uncertain $m=0$ identifications. If a frequency present appears near where we might expect a rotationally split multiplet component but does not exceed the 0.34\% significance threshold, we include it here but mark it with an asterisk ($^{*}$).
\label{tab:splitmodes}} 
\tablenum{2}
\tablehead{
\colhead{ID} & 
\colhead{\hspace{.1cm}$\ell$}\hspace{.1cm} & \colhead{\hspace{.2cm}$m$}\hspace{-.05cm} & \colhead{\hspace{.7cm}Period (s)}\hspace{.7cm} &
\colhead{\hspace{.2cm}Frequency ($\mu$Hz)}\hspace{.2cm} &
\colhead{Amp. (\%)} &
\colhead{Splitting ($\mu$Hz)}}
\tablecolumns{7}
\startdata
$f_{1a}$	&	1	&	$-$1	&	5135.1	$\pm$	0.8 	&	194.74	$\pm$	0.03	&	0.47	& \\
$f_{1b}$	&	1	&	0	&	{\bf 4756.5}	$\pm$	0.7 	&	210.24	$\pm$	0.03	&	0.47	&	15.50	$\pm$	0.06	\\
$f_{1c}$	&	1	&	1	&	4428.5	$\pm$	0.5 	&	225.812	$\pm$	0.028	&	0.53	&	15.57	$\pm$	0.06	\\
\hline
$f_{2a}$	&	1	&	$-$1	&	4259.1	$\pm$	0.5 	&	234.79	$\pm$	0.03	&	0.50	&				\\
$f_{2b}$	&	1	&	0	&	{\bf 4028.83}	$\pm$	0.21	&	248.211	$\pm$	0.013	&	1.19	&	13.42	$\pm$	0.04	\\
$f_{2c}^{*}$	&	1	&	1	&	3815.8	$\pm$	0.7	&	262.07	$\pm$	0.05	&	0.32	&	13.85	$\pm$	0.06	\\
\hline
$f_{3a}$	&	1	&	$-$1	&	3365.56	$\pm$	0.14	&	297.127	$\pm$	0.012	&	1.25	&				\\
$f_{3b}^{*}$	&	1	&	0$^{\dagger}$	&	{\bf 3226.1}	$\pm$	0.6 	&	309.98	$\pm$	0.06	&	0.24	&	12.85	$\pm$	0.07	\\
\hline
$f_{4a}$	&	1	&	0$^{\dagger}$	&	{\bf 2658.51}	$\pm$	0.28 	&	376.15	$\pm$	0.04	&	0.35	& \\
$f_{4b}^{*}$	&	1	&	+1	&	2569.0	$\pm$	0.4 	&	389.26	$\pm$	0.06	&	0.26	&	13.10	$\pm$	0.10	\\
\hline
$f_{5a}$	&	1	&	$-$1	&	2447.86	$\pm$	0.18	&	408.52	$\pm$	0.03	&	0.46	&				\\
$f_{5b}$  &   1   &     0     &   {\bf 2367.4} $\pm$ 0.21      &                            &           &               \\
$f_{5c}^{*}$	&	1	&	1	&	2292.0	$\pm$	0.3 	&	436.29	$\pm$	0.06	&	0.26	&	2\,$\times$\,(13.89	$\pm$	0.09)	\\
\hline
$f_{6a}$	&	2	&	$-$2	&	6729.0	$\pm$	1.8 	&	148.61	$\pm$	0.04	&	0.38	&				\\
$f_{6b}$	&	2	&	$-$1	&	5896.8	$\pm$	1.1 	&	169.58	$\pm$	0.03	&	0.48	&	20.97	$\pm$	0.07	\\
$f_{6c}$	&	2	&	0	&	{\bf 5253.2}	$\pm$	0.19	&	190.359	$\pm$	0.007	&	2.15	&	20.78	$\pm$	0.04	\\
$f_{6d}^{*}$	&	2	&	2	&	4270.8	$\pm$	0.8 	&	234.15	$\pm$	0.05	&	0.32	&	2\,$\times$\,(21.89	$\pm$	0.05)	\\
\hline
$f_{7a}$	&	2	&	$-$2	&	6295.4	$\pm$	1.0 	&	158.846	$\pm$	0.026	&	0.58	&				\\
$f_{7b}$	&	2	&	$-$1	&	5550.6	$\pm$	0.8 	&	180.161	$\pm$	0.025	&	0.61	&	21.32	$\pm$	0.05	\\
$f_{7c}$	&	2	&	0	&	{\bf 4959.6}	$\pm$	1.0 	&	201.63	$\pm$	0.04	&	0.37	&	21.46	$\pm$	0.07	\\
$f_{7d}^{*}$	&	2	&	1	&	4473.1	$\pm$	0.9 	&	223.56	$\pm$	0.05	&	0.32	&	21.93	$\pm$	0.09	\\
$f_{7e}$	&	2	&	2	&	4083.5	$\pm$	0.7 	&	244.89	$\pm$	0.04	&	0.37	&	21.33	$\pm$	0.09	\\
\hline
$f_{8a}$	&	2	&	$-$1	&	5814.8	$\pm$	0.7 	&	171.975	$\pm$	0.021	&	0.71	&				\\
$f_{8b}$	&	2	&	0$^{\dagger}$	&	{\bf 5198.3}	$\pm$	0.8 	&	192.370	$\pm$	0.029	&	0.53	&	20.39	$\pm$	0.05	\\
$f_{8c}$	&	2	&	+1	&	4628.8	$\pm$	0.5 	&	216.039	$\pm$	0.022	&	0.68	&	23.67	$\pm$	0.05	\\
\enddata
\end{deluxetable*}

\subsection{Possible Multiplet Mode Identification} \label{sec:modes}
We find evidence for patterns in the observed periodogram that suggest common frequency spacings that could arise from rotational splittings if the ELM white dwarf is rotating at roughly 10\,hr, which would cause $\ell=1$ mode splittings of $\approx13.7$\,\muhz\ and $\ell=2$ mode splittings of $\approx21.5$\,\muhz\ (e.g., \citealt{2021RvMP...93a5001A}).

In total, we identify eight mode splittings in the GD\,278 \emph{TESS} periodogram: five that are consistent with $\ell=1$ modes and three $\ell=2$ modes. In Table~\ref{tab:splitmodes}, we list 24 individual observed pulsation periods that make up eight groups, one for each multiplet. Also listed in Table~\ref{tab:splitmodes} are the assumed $\ell$ and $m$ numbers, periods, frequencies, amplitudes, and the amount adjacent modes within a group are split by. We observe weighted mean rotational splittings of roughly 14.0\,\muhz\ for the  $\ell=1$ modes and 21.5\,\muhz\ for the $\ell=2$ modes, both consistent with an overall rotation rate of roughly 10\,hr.

The six modes with an asterisk have amplitudes less than the significance threshold; although these amplitudes lie below the significance threshold, we relax this threshold based on their location near predictions from common frequency splittings (e.g., \citealt{1991ApJ...378..326W, 2010ApJ...720L.159D}). 

\begin{figure}[t]
\epsscale{1.15}
\plotone{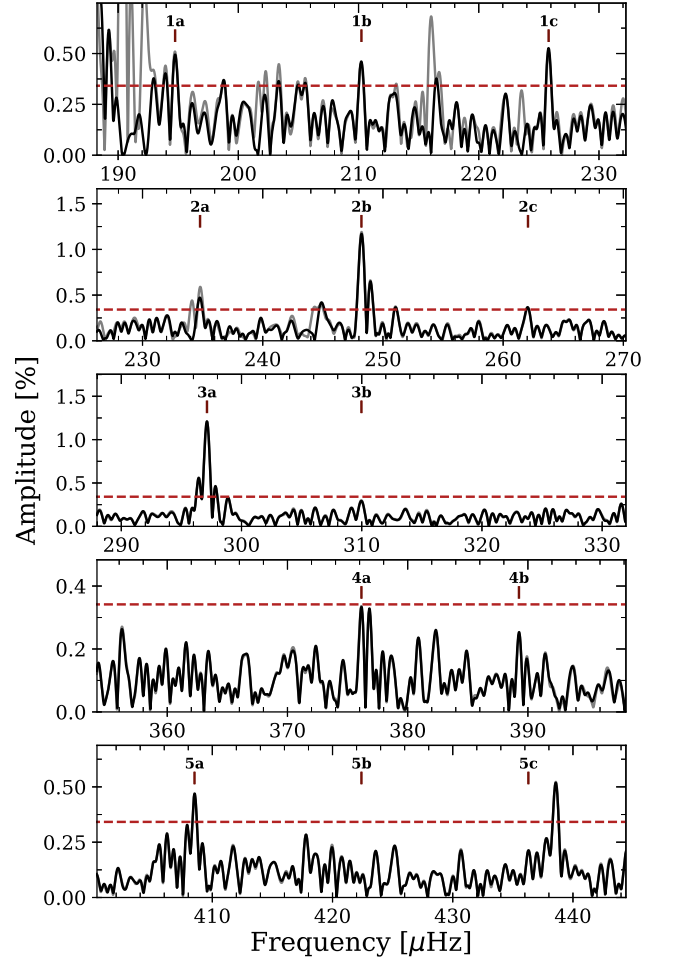}
\caption{Possible evidence for five $\ell = 1$ mode splittings, identified within the GD\,278 \emph{TESS} periodogram and listed in Table \ref{tab:splitmodes}. Red tick marks indicate the frequencies that we propose form rotationally split multiplets, consistent with an overall 10\,hr rotation rate. For each mode we show in grey the prewhitened \emph{TESS} light curve, subtracted of all frequencies listed in Table \ref{tab:freq_sol} other than those indicated by the red points, and show the resulting periodogram in black. The red dashed lines mark the 0.34\% significance threshold. \label{fig:ell_1}}
\end{figure}

We show an illustration of the five $\ell = 1$ splittings in Figure~\ref{fig:ell_1}. The location of the modes are indicated by the red tick marks, while the red dashed line represents the 0.1\% False Alarm Probability threshold at 0.34\% amplitude. In grey, we plot the periodogram of the original \emph{TESS} dataset. Over this periodogram, we plot in black a periodogram of the \emph{TESS} data, after prewhitening by all periods listed in Table~\ref{tab:splitmodes}, not including the group of split modes being shown in a given subplot. In some cases ($f_3$ and $f_4$) we only identify two components, so it is not possible to unambiguously determine which of the two components corresponds to the central ($m=0$) mode. There is therefore ambiguity on which modes should be considered the $m=0$ mode used for the asteroseismic analysis in Section~\ref{sec:modeling}.

\begin{figure}[t]
\epsscale{1.15}
\plotone{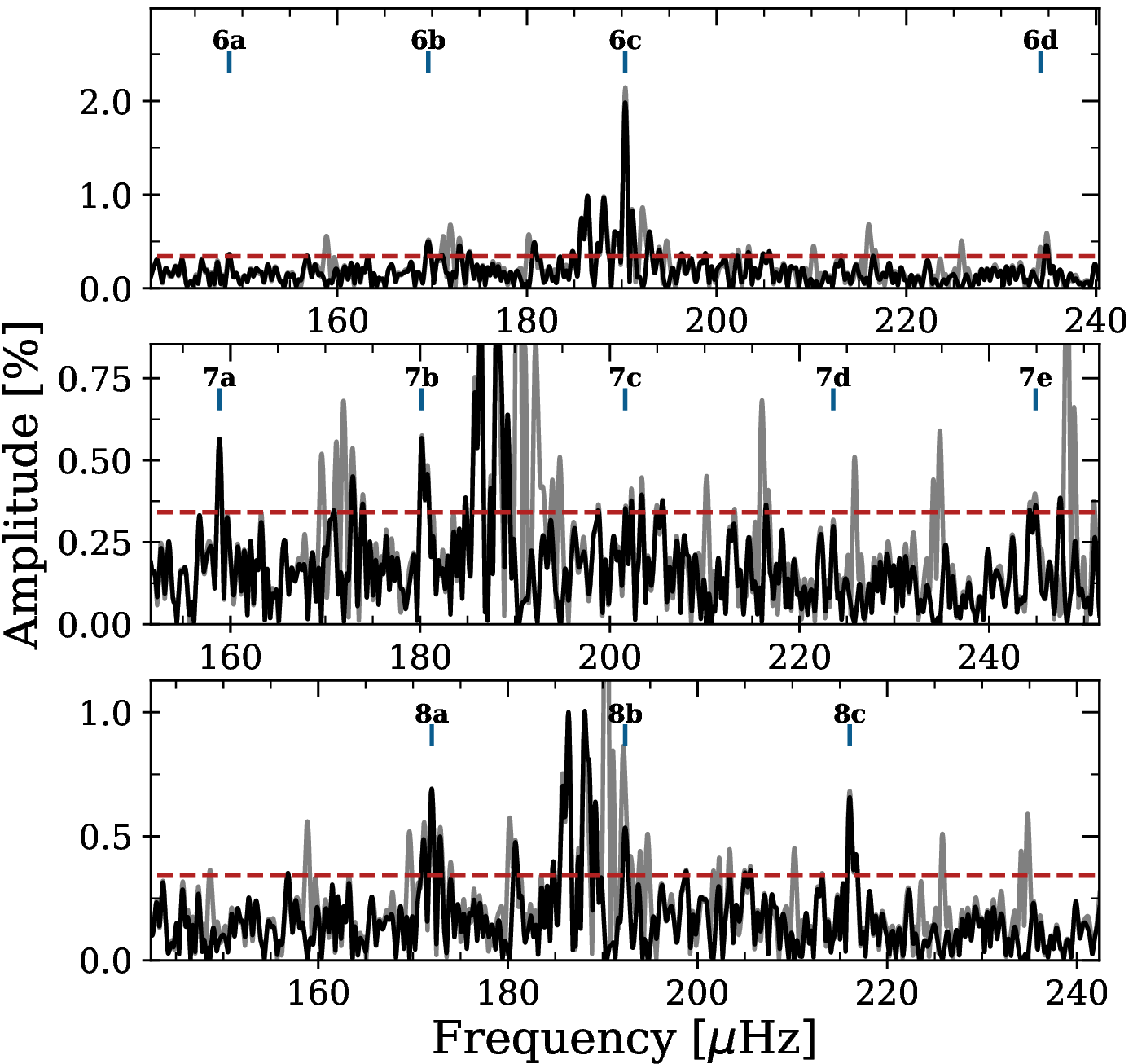}
\caption{Similar to Figure \ref{fig:ell_1}, except for the three $\ell = 2$ mode splittings identified identified within the GD\,278 \emph{TESS} periodogram and listed in Table~\ref{tab:splitmodes}. Blue tick marks indicate the frequencies that we propose form the rotationally split multiplets, also consistent with 10\,hr rotation.   \label{fig:ell_2}}
\end{figure}

We repeat this illustration for the three identified $\ell = 2$ modes in Figure~\ref{fig:ell_2}, this time indicating the location of the modes by blue tick marks. One of the $\ell=2$ modes ($f_8$) only has three identified components, which again leads to ambiguity in determining which of the three components corresponds to the $m=0$ mode.

\section{Asteroseismic Modeling} \label{sec:modeling}

The $m=0$ modes we identify in Table~\ref{tab:splitmodes} serve as the baseline for our asteroseismic investigation into GD\,278, offering us five likely $\ell=1$ modes and three likely $\ell=2$ modes observed in this pulsating white dwarf. For completeness, our full asteroseismic analysis explores solutions using all 12 different possible combinations of identified $m=0$ modes from the ambiguous $\ell=1$ modes of $f_3$ and $f_4$, as well as the ambiguous $\ell=2$ mode of $f_8$.

\subsection{Initial Expectations from MESA Modeling} \label{subsec:MESA}

As an initial pass at guiding expectations, we generated a model of the ELM white dwarf GD 278 using the stellar evolution code \texttt{MESA} version r12115 \citep{2011ApJS..192....3P,2013ApJS..208....4P,2015ApJS..220...15P,2018ApJS..234...34P,2019ApJS..243...10P}. Our input model to generate a 0.19\,\msun\ ELM white dwarf followed the inlists first described by \citet{2016A&A...595A..35I}, which evolve a close binary and include element diffusion and rotational mixing. In short, we start with a 1.1\,\msun, $Z$\,=\,0.0142 metallicity ZAMS progenitor in a 2.90-day binary with a 1.4\,\msun\ companion and evolve the system through mass-loss until the remnant is a 0.19\,\msun\ ELM white dwarf; further details are discussed in \citet{2016A&A...595A..35I}.

We evolved this ELM white dwarf model through its many diffusion-induced CNO flashes (e.g., \citealt{2001MNRAS.323..471A,2007MNRAS.382..779P, 2013A&A...557A..19A}) until it arrived at the final cooling track, and paused the evolution at 9180 K. Here our 0.1904\,\msun\ model has a surface gravity of \logg\ = 6.644, a hydrogen-layer mass of $M_{H} = 8.2 \times 10^{-4} M_\star$, a radius of $R_\star = 0.03443 R_\odot$, and a luminosity of $L_{\star} =7.56 \times 10^{-3} L_{\odot}$. We artificially smoothed the \bvf\ frequency from this static model with a built-in weighted smoothing in MESA using eight cells (rather than the default two cells), in order to avoid any numerical noise dominating our resultant pulsation frequencies.

We took the output of this smoothed \texttt{MESA} model and pulsated it using \texttt{GYRE} \citep{2013MNRAS.435.3406T}, to generate an output set of expected non-radial pulsation modes. We did not attempt to generate a grid of models to explore to perform a full asteroseismic fit, but first sought to understand the possible radial order of the observed modes in GD\,278 from the {\em TESS} observations. We compare the observed periods to the theoretical \texttt{MESA} predictions in Table~\ref{tab:MESAperiods}, and especially note that these are not fits but rather the output of a model selected to match the spectroscopy of GD\,278. As a reminder, the atmospheric parameters from spectroscopy by \citet{2020ApJ...889...49B} include the 3D corrections of \citet{2015ApJ...809..148T} and find ${T}_{\mathrm{eff}} = 9230 \pm 100$\,K and $\log{g} = 6.627 \pm 0.056$, which corresponds to a mass of $0.191\pm0.013$\,$M_{\odot}$.

\begin{deluxetable}{cccccc}[t!]
\tablecaption{Properties of the theoretical mode periods ($\Pi^T$) of a representative \texttt{MESA} model we obtained for GD\,278 defined by the spectroscopic parameters from \citet{2020ApJ...889...49B}: $M = 0.1904$\,\msun, $M_{H} = 8.2 \times 10^{-4} M_\star$, \teff = 9180 K, $L_{\star} =7.56 \times 10^{-3} L_{\odot}$, \logg = 6.644, and $R_\star = 0.03443 R_\odot$. The observed periods identified in  Table~\ref{tab:splitmodes} are also included ($\Pi^O$). This model has $1/\chi^2=0.0024$ (see Equation~\ref{eq:1}). \label{tab:MESAperiods}}
\tablenum{3}
\tablehead{
\colhead{ID} & \colhead{$\Pi^O$ (s)} & \colhead{\hspace{.65cm}$\Pi^T$ (s)}\hspace{0.65cm} & \colhead{\hspace{0.35cm}$\ell$}\hspace{.35cm} & \colhead{\hspace{.35cm}$k$}\hspace{.35cm} & \colhead{$\delta \Pi$ (s)} }
\tablecolumns{6}
\startdata
$f_{1}$ & 4756.5 &	4761.8 &	1 &	56 &	$-$5.3 \\
$f_{2}$ & 4028.83 &	4020.8 &	1 &	47 &	8.0 \\	
$f_{3}$ & 3226.1 &	3204.6 &	1 &	37 &	21.5  \\	
$f_{4}$ & 2658.51 &	2637.7 &	1 &	30 &	20.8 \\	
$f_{5}$ & 2367.4 &	2395.0 &	1 &	27 &	$-$27.6 \\	
$f_{6}$ & 5253.2 &	5270.2 &	2 &	108 &	$-$17.0 \\	
$f_{7}$ & 4959.6 &	4932.8 &	2 &	101 &	26.8 \\	
$f_{8}$ & 5198.3 &	5221.8 &	2 &	107 &	$-$23.5
\enddata
\end{deluxetable}

This \texttt{MESA} model features an overall $\ell=1$ asymptotic mean period spacing of roughly 83\,s, and an $\ell=2$ mean period spacing of roughly 48\,s. The observed period range in Table~\ref{tab:MESAperiods} implies that the $\ell=1$ modes have radial orders ranging from $27 < k < 56$, with $\ell=2$ modes of especially high radial order, ranging from $101 < k < 108$.

Our stellar model includes rotation, with the initial rotation velocity set such that the star is synchronized with the initial orbital period. Moreover, we take into account the effect of tides and spin-orbit coupling. Given this, we note that this final \texttt{MESA} model has a 11.6-hr surface rotation period. Interestingly, the model is rotating marginally differentially, a consequence imprinted in the CNO flashing episodes (see Section~4.1.1 in \citealt{2016A&A...595A..35I}). If the rotational splittings we identify in Section~\ref{sec:modes} are correct, this model predicts rotation at almost exactly the same rate as observed in GD\,278.

We can also estimate the expected distance to this white dwarf by comparing the luminosity of the model ($L_\star = 0.00756$\,\lsun) to the \emph{Gaia} apparent magnitude ($G$=14.88\,mag). Using the 3D reddening map of \citep{2019ApJ...887...93G}, we find that the effects of reddening do not become an issue at the coordinate of GD\,278 until $>$200 pc, so we assume no reddening coefficient. We use the bolometric correction of \citep{2016ApJ...823..102C} for an object with \teff = $9250$\,K and \logg\ = 6.5. We obtain a seismic distance of 104.4\,pc. This is significantly discrepant with the parallax distance determined from {\em Gaia} eDR3 of $151.19\pm0.78$\,pc \citep{2021AJ....161..147B}, suggesting our model is underluminous compared to that needed to explain the measured {\em Gaia} parallax.

\subsection{Asteroseismic Fits with LPCODE} \label{subsec:LaPlata}

We further analyzed the identified periods observed in GD\,278 using the low-mass white dwarf models of \citet{2018A&A...620A.196C}. In short, the models expand upon the ELM white dwarf models described in \citet{2013A&A...557A..19A} generated using the {\tt LPCODE} stellar evolution code. This code computes in detail the complete evolutionary stages that lead to the white dwarf formation, allowing the study of the white dwarf evolution consistently with the predictions of the evolutionary history of the progenitors \citep[see,][for details]{2005A&A...435..631A,2013A&A...557A..19A, 2015A&A...576A...9A}. For ELM white dwarfs, {\tt LPCODE} computes initial models by mimicking the binary evolution of progenitor stars \citep{2013A&A...557A..19A}. 
Adiabatic pulsation periods for non-radial dipole ($\ell = 1$) and quadrupole ($\ell = 2$) $g$-modes were computed employing the adiabatic version of the {\tt LP-PUL} pulsation code \citep[see,][for details]{2006A&A...454..863C}. The models of low-mass white dwarfs we employ in this work include not just canonically thick hydrogen layers, but also include new evolutionary sequences of thinner envelopes \citep{2018A&A...620A.196C}. 

We searched a grid of ELM white dwarf models for the best-fit solution in a broad range of effective temperatures ($8500-9800$\,K), overall mass ($0.15-0.44$\,\msun), and hydrogen envelope thicknesses in the interval $-5.8 \lesssim \log(M_{\rm H}/M_{\star}) \lesssim -1.7$ (depending on the stellar mass). Eventually we restricted our attention to the region within roughly 4$\sigma$ of the overall mass determined from the spectroscopic parameters ($0.15-0.24$\,\msun).

Using this grid of models, we searched the 12 different periods solutions that arise from the ambiguity of the $f_3$, $f_4$, and $f_8$ splittings (Section~\ref{sec:modes}), fitting only the $m=0$ modes. We explored best-fit models for each of the 12 period solutions, both restricted to the narrow range around the spectroscopic solution and across the broader range of overall stellar mass and effective temperature. Our best-fit models were optimized to minimize a quality function ($\chi^2$) that considers the differences between the observed pulsation periods, $\Pi^O$, and theoretical pulsation periods, $\Pi^T$, and is defined as 

\begin{equation} \label{eq:1}
\chi^2 \left ( M_\star, {T}_{\mathrm{eff}}, M_H \right ) = \frac{1}{N} \sum_{i=1}^{N} \left ( \Pi^O_i - \Pi^T_i \right )^2
\end{equation}

\noindent
where $M_\star$ is the stellar mass, \teff \ is the effective temperature, $M_H$ is the mass fraction of the hydrogen envelope, and $N$ is the number of observed pulsation periods. For this procedure, we fix the values of $\ell$ at the outset, according to the identifications in Table~\ref{tab:splitmodes}.

Unfortunately, there is not a global minimum preferred fit within the 1$\sigma$ bounds provided by the spectroscopy of \citet{2020ApJ...889...49B}. For all 12 period solutions, the ELM white dwarf model with the formal best fit to the observed $m=0$ pulsation periods occurs with a stellar mass of 0.239\,\msun, an effective temperature between $9680-9710$\,K, and a hydrogen-layer mass of roughly $M_{H} = 3.6 \times 10^{-3} M_\star$. 

\begin{figure}[t] 
\hspace*{-0.5cm}
\epsscale{1.2}
\plotone{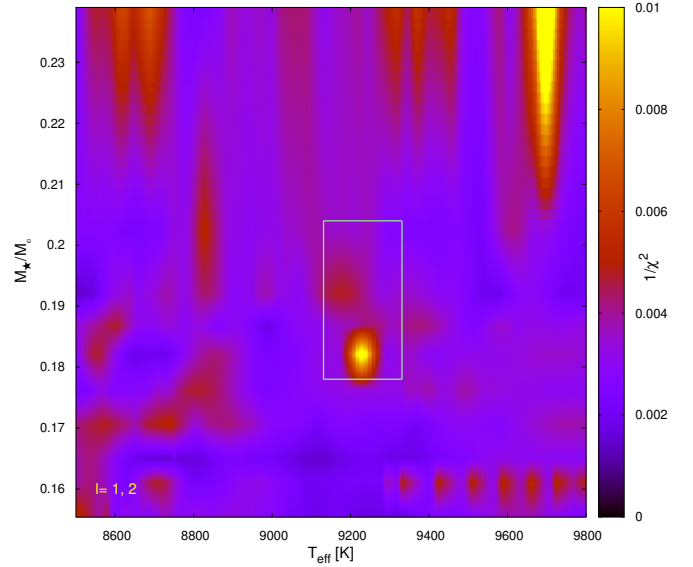}
\caption{Map of the inverse of the quality function ($1/\chi^2$) as a function of the stellar mass ($M_\star$) and effective temperature (\teff), considering the eight $m=0$ modes identified in bold in Table~\ref{tab:splitmodes}. Each point on the map corresponds to a hydrogen envelope mass ($M_H$) that maximizes the value of $1/\chi^2$. The green box represents the GD\,278 spectroscopic mass and effective temperature, and the corresponding $1\sigma$ uncertainties, from \citet{2020ApJ...889...49B}.
\label{fig:grid1}}
\end{figure}

As all 12 period solutions return similar (though not identical) quality maps, we restrict further discussion to only the solution defined explicitly in Table~\ref{tab:splitmodes}. The quality of the fit to the observed periods for this solution across all effective temperatures and stellar masses is shown in Figure~\ref{fig:grid1}. Each point ($M_{\star}$, $T_{\rm eff}$) in the map corresponds to a hydrogen-layer mass ($M_{\rm H}$/$M_{\star}$) that maximizes the value of $1/\chi^2$ for that stellar mass and effective temperature. The green box represents the 1$\sigma$ bounds of the mass and effective temperature obtained from spectroscopy. For this model, the formal best (global) solution has $1/\chi^2$ = 0.0154, with an overall mass of 0.239\,\msun, a hydrogen-layer mass of $M_{H} = 3.55 \times 10^{-3} M_\star$, and an effective temperature of 9698\,K. Notably, this model is $>$3$\sigma$ in tension with both the observed spectroscopic effective temperature and mass.

Similar to Section\,\ref{subsec:MESA}, we estimate the distance to this white dwarf by comparing the seismic luminosity of the global best-fit model ($L_\star = 0.00645$\,\lsun) to the \emph{Gaia} apparent magnitude. We obtain a seismic distance of 96.4\,pc for this model. Again, this value is significantly discrepant with the $151.19\pm0.78$\,pc parallax distance \citep{2021AJ....161..147B}.

If we restrict our fits and require them to be within a 1$\sigma$ box consistent with the spectroscopic parameters, the best-fit model occurs at an overall mass of 0.1822\,\msun\ (with \logg\ = 6.727) and an effective temperature of 9226\,K, with an extremely thin hydrogen-layer mass of $M_{H} = 3.08 \times 10^{-6} M_\star$, more than three orders of magnitude thinner than the global best-fit model, as well as the exploratory \texttt{MESA} model in Section~\ref{subsec:MESA}. This restricted solution has a goodness of fit value of $1/\chi^2$ = 0.0111. Our period solution and radial orders of this restricted best-fit are described in Table~\ref{tab:LaPlata}. The observed period range in Table~\ref{tab:LaPlata} implies that the $\ell=1$ modes have radial orders of $22 < k < 46$, with $\ell=2$ modes ranging from $83 < k < 88$. For this model, with $L_\star = 0.00612$\,\lsun, we find a seismic distance of 93.9\,pc. 

\begin{deluxetable}{cccccc}[t!]
\tablecaption{Properties of the best-fit asteroseismic model for GD\,278 using the {\tt LPCODE} evolution code, restricted to within 1$\sigma$ of the spectroscopic parameters, where all modes are assumed to be the $m=0$ components identified in Table\,\ref{tab:splitmodes}. The characteristics of this best-fit model are as follows: $M_{\star} = 0.182 \ M_{\odot}$, $M_{H} = 3.1 \times 10^{-6} M_\star$, \teff = 9226\,K, $L_{\star} = 6.12 \times 10^{-3} L_{\odot}$, \logg = 6.727, $R_\star = 0.0306 R_\odot$, and $1/\chi^2$ = 0.0111.  \label{tab:LaPlata}}
\tablenum{4}
\tablehead{
\colhead{ID} & \colhead{$\Pi^O$ (s)} & \colhead{\hspace{.65cm}$\Pi^T$ (s)}\hspace{0.65cm} & \colhead{\hspace{0.35cm}$\ell$}\hspace{.35cm} & \colhead{\hspace{.35cm}$k$}\hspace{.35cm} & \colhead{$\delta \Pi$ (s)} }
\tablecolumns{6}
\startdata
$f_{1}$ & 4756.5 &	4760.13 &	1 &	46 &	$-$3.63 \\
$f_{2}$ & 4028.83 &	4043.10 &	1 &	38 &	$-$14.27 \\	
$f_{3}$ & 3226.1 &	3238.14 &	1 &	31 &	$-$12.04  \\	
$f_{4}$ & 2658.51 &	2644.92 &	1 &	25 &	13.59 \\	
$f_{5}$ & 2367.4 &	2355.23 &	1 &	22 &	12.17 \\	
$f_{6}$ & 5253.2 &	5253.85 &	2 &	88 &	$-$0.65 \\	
$f_{7}$ & 4959.6 &	4956.15 &	2 &	83 &	3.45 \\	
$f_{8}$ & 5198.3 &	5194.41 &	2 &	87 &	3.89
\enddata
\end{deluxetable}

This thin-hydrogen-layer solution at 0.182\,\msun\ is the most commonly occurring restricted best-fit (in eight of 12 cases), and in all eight cases implies an $\ell=1$ mean period spacing of roughly 100.0\,s. In the other four cases, the best restricted fit (within 1$\sigma$ of the spectroscopic parameters) occurs with models featuring an overall mass of 0.192\,\msun, effective temperatures ranging from $9340-9360$\,K, and a thicker hydrogen-layer mass of $M_{H} = 2.38 \times 10^{-4} M_\star$. With the slightly higher overall mass, the $\ell=1$ mean period spacing for these models is roughly 91.4\,s. The goodness-of-fit parameter for the 0.192\,\msun\ solution is worse than the 0.182\,\msun\ solution, with $1/\chi^2$ = 0.0093. However, it is most proximate to the spectroscopic values without invoking an artificially stripped hydrogen-layer mass that is much thinner than what is expected from stellar evolution \citep{2007MNRAS.382..779P,2010ApJ...718..441S}.

Given that our absolute best-fit model across all parameters disagrees significantly with the spectroscopic parameters, we are reluctant to use our asteroseismic analysis to adopt an overall stellar model. This could implicate our mode identifications and our assumption of a roughly 10-hr rotation period in Section~\ref{sec:modes} as being systematically incorrect. We note that some of our splittings within the same multiplet are inconsistent; for example, the three modes in $f_2$ have frequency spacings that are formally $>$$4\sigma$ discrepant. However, the vast majority of rotational multiplets identified in Table~\ref{tab:splitmodes} have frequency separations consistent to within $2\sigma$, and the implied rotation rate agrees for both the $\ell=1$ and $\ell=2$ modes, independently.

It is also possible that even if we have correctly identified all eight $m=0$ modes present, the relatively long periods in GD\,278 have such high radial order that none of the modes are particularly sensitive to interior structure transitions. These long-period modes may not lend themselves to a strongly favored, unique global solution, as seen in some higher-mass DAVs (e.g., \citealt{2017A&A...598A.109G}).

\subsection{Search for a Mean Period Spacing}

Based on some of the ambiguities in the asteroseismic analysis, we sought to constrain at least the mean density of the ELM white dwarf GD\,278 by trying to measure and then analyze the observed mean period spacing, which approximates the overall mean density and thus overall mass of the star using the $g$-mode pulsations present \citep{2010A&ARv..18..471A}.

First, we attempted to search for a roughly constant period spacing in our observations. Since we cannot know the radial order ($k$) of the modes a priori, we first searched for any sign in the data for recurring period spacings. The long-period pulsations present ($>2300$\,s) imply relatively high radial orders ($k>20$), so the period spacing between modes should be asymptotic and nearly constant \citep{1980ApJS...43..469T}.

\begin{figure}[t]
\epsscale{1.15}
\plotone{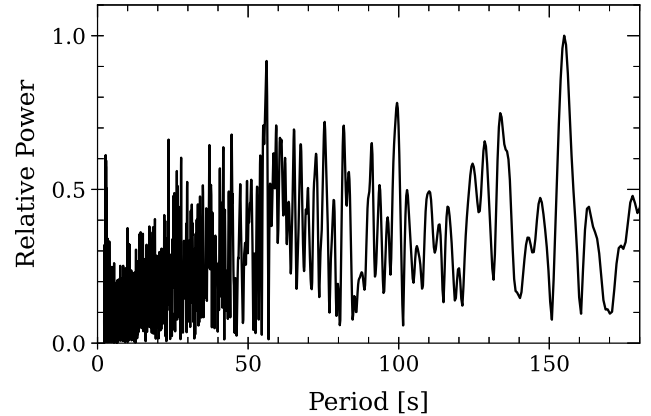}
\caption{Period transform of the GD\,278 Lomb-Scargle periodogram shown in Figure~\ref{fig:ft}. Unfortunately, we cannot see any clear recurrent period spacings in the periodogram itself. \label{fig:pt}}
\end{figure}

\begin{figure}[t]
\epsscale{1.2}
\plotone{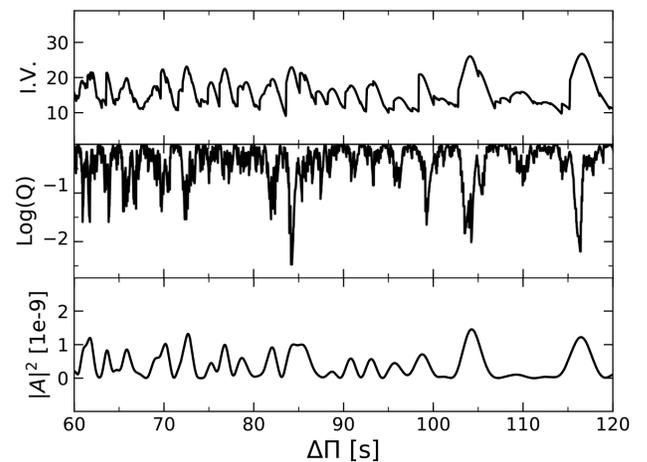}
\caption{We searched for a constant period spacing among all detected significant pulsation frequencies listed in Table~\ref{tab:freq_sol} using three tests: an Inverse Variance test (top panel), a Kolmogorov-Smirnov test (middle panel), and a Fourier Transform test (bottom panel). The strongest signals emerge at roughly 104\,s, 84\,s, 116\,s, and 72.5\,s, and are put in context in Figure~\ref{fig:spacing}. \label{fig:3tests}}
\end{figure}

We initially searched for patterns in the period spacing in the data itself, by taking a period transform of the period transform; in short, we inverted the Lomb-Scargle periodogram in Figure~\ref{fig:ft} so that it was in period and not frequency, and then took the period transform of this period transform (e.g., \citealt{1991ApJ...378..326W}) within the range of all pulsation power ($2000-8000$\,s). This is shown in Figure~\ref{fig:pt}. Unfortunately this did not reveal any dominant peaks, so was not valuable for determining recurrent period spacings in the data itself.

As a next step, we searched for a constant period spacing using three additional tests (e.g., \citealt{2017A&A...607A..33C}). In short, we ignored our mode identifications and searched all significant frequencies given in Table~\ref{tab:freq_sol} using a Kolmogorov-Smirnov test (K-S test; \citealt{1988IAUS..123..329K}), an Inverse Variance test (I-V test; \citealt{1994MNRAS.270..222O}), and the Fourier Transform test (F-T test; \citealt{1997MNRAS.286..303H}). The results of the three tests are shown in Figure~\ref{fig:3tests}. There are some suggestive indications of constant period spacings at roughly 72.5\,s, 84\,s, 104\,s, and 116\,s; most prominently at 84\,s.

\begin{figure}[t]
\hspace*{-0.5cm}
\epsscale{1.2}
\plotone{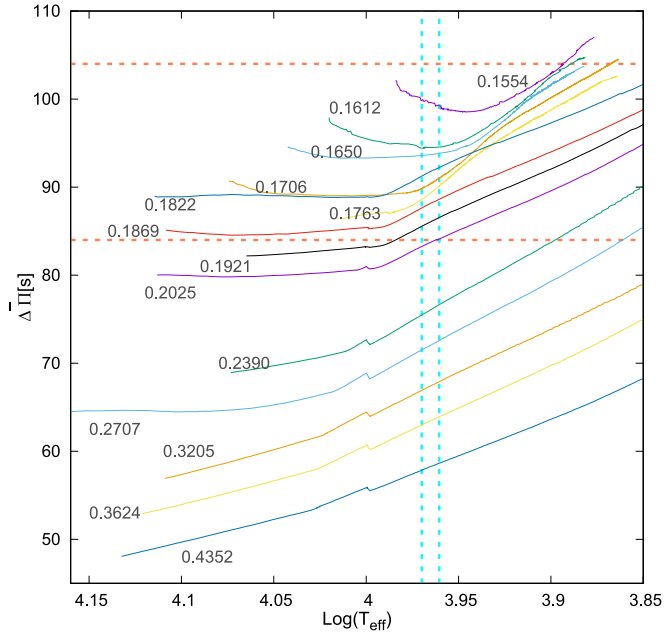}
\caption{Average of the computed dipole period spacings  for $\ell=1$ modes as a function of effective temperature for various cooling models of ELM white dwarfs \citep{2018A&A...620A.196C} with masses between $0.15-0.44$\,\msun\ and canonically thick hydrogen envelopes. Plotted over these models are the two possible values of the constant period spacing for GD\,278 found from our analysis in Figure~\ref{fig:3tests} (roughly 84\,s and 104\,s, horizontal lines), as well as bounds from the spectroscopic effective temperature and uncertainties (vertical cyan lines). For white dwarfs $<$0.2\,\msun\ we expect mean period spacings $\overline{\Delta \Pi} > 80$\,s.
\label{fig:spacing}}
\end{figure}

To put these mean period spacings into context, we show in Figure~\ref{fig:spacing} a series of expected period spacings as a function of effective temperature for a suite of ELM white dwarf models with masses between $0.15-0.44$\,\msun\ (described in Section~\ref{subsec:LaPlata}). All models have the thickest possible (canonical) hydrogen-layer masses: sequences with $M_{\star} < 0.18 \ M_{\odot}$ are characterized by $M_{H} \approx 10^{-2} M_\star$ (since these model ELM white dwarfs do not undergo CNO flashes), while those with $M_{\star} > 0.18 \ M_{\odot}$ have $M_{H} < 2 \times 10^{-3} M_\star$. The effective temperature constraints from spectroscopy are shown as vertical cyan lines.

Figure~\ref{fig:spacing} shows that the 116\,s possible period spacing is not within the range expected for $\ell=1$ modes of ELM white dwarfs.  The roughly 104\,s, 84\,s, and 72.5\,s period spacings would imply values of the overall stellar masses lower than 0.14\,$M_{\odot}$, roughly 0.20\,$M_{\odot}$, and roughly 0.24\,$M_{\odot}$, respectively. The two $\ell=1$ mean period spacings in Figure~\ref{fig:3tests} at 104\,s and 84\,s are marked with horizontal orange lines. We also repeated the exercise shown in Figure~\ref{fig:3tests} on the smaller list of eight likely $m=0$ modes from Table~\ref{tab:splitmodes}, which includes a mix of $\ell=1$ and $\ell=2$ modes. We do not visualize this result but these tests show slight preference for possible mean period spacings at roughly 62\,s and 72\,s, which fall at an unexpected range based on previous spectroscopy of GD\,278, so we do not discuss those tests further.

The strongest response to the K-S test in Figure~\ref{fig:3tests} occurs at roughly 84\,s, which is very similar to the $\ell=1$ mean period spacing of roughly 82\,s that we expect from our \texttt{MESA} modeling described in Section~\ref{subsec:MESA} for a white dwarf of the given spectroscopic parameters and a canonically thick hydrogen layer. However, the ambiguity of this signal and the mode identification presented in Table~\ref{tab:splitmodes} precludes closing the chapter on asteroseismic investigation of GD\,278 until we can find a significantly best-fit solution. Improved fits are likely to come from utilizing the precise parallax distance from {\em Gaia}, which we leave to future work.

\section{Discussion and Conclusions} \label{sec:discussion}
We discovered GD\,278 as part of a search of the \cite{2019MNRAS.482.4570G} \emph{Gaia} DR2 catalog of white dwarfs for pulsating ELM white dwarfs. It was selected for its position on the \emph{Gaia} color-magnitude diagram relative to other known pulsators and on its intrinsic variability, which we inferred from its \emph{Gaia} $G$-band flux uncertainty. This method for selecting candidate variable stars was described further by \cite{2021ApJ...912..125G}, who used the method to confirm 32 new DAVs and one new pulsating ELM white dwarf. 

GD\,278 stood out as a likely variable candidate and we confirmed its pulsations from McDonald Observatory in 2019, before it was announced as a bona fide ELM white dwarf in a 4.61-hr single-lined binary by \citet{2020ApJ...889...49B}, selected from an existing catalog of bright white dwarfs \citep{2017MNRAS.472.4173R}. Our ground-based discovery of pulsations allowed us to obtain 2-min cadence observations with {\em TESS}, obtaining the first space-based photometry of a pulsating ELM white dwarf. This revealed the richest set of oscillation periods ever obtained for an ELM white dwarf.

However, mode identification of the frequencies in Table~\ref{tab:splitmodes} was a particular challenge, with rotational splitting exceeding the frequency differences between adjacent radial overtones. One method for identifying the spherical degree ($\ell$) is to look for triplets or quintuplets of modes with frequencies consecutively split by equal amounts. For GD\,278, we identified five possible $\ell=1$ splittings and three possible $\ell=2$ splittings, many overlapping with one another. The $\ell=1$ modes we observe have a mean frequency separation of $\approx$$14.0$\,\muhz, and the $\ell=2$ modes are separated from one another by $\approx$$21.5$\,\muhz; both are consistent with an overall rotation period of roughly 10\,hr. This is the first direct estimate of the rotation rate of an ELM white dwarf. The role of rotation is likely vital for understanding the frequent occurrence of heavy metals in the atmospheres of ELM white dwarfs \citep{2014MNRAS.444.1674H,2016A&A...595A..35I,2016A&A...595L..12I}.

We recover the $4.6092\pm0.0048$\,hr orbital period from the spectroscopy of \citet{2020ApJ...889...49B} in the \emph{TESS} photometry, finding a marginal but significant variation at $4.6142\pm0.0022$\,hr. The 4.61-hr orbital period of GD\,278 is very close to the $4.5912\pm0.0012$\,hr orbital period of SDSS\,J184037.78+642312.3, which was the first ELM white dwarf discovered to exhibit pulsations \citep{2012ApJ...750L..28H}. That work notes that pulsating ELM white dwarf with a rotation period of 4.6\,hr would have $\ell=1$ modes split into components separated by 30 \muhz, but a 4.6\,hr orbital period is likely not short enough for the white dwarf to have synchronized with the orbital period through tidal dissipation \citep{2013MNRAS.430..274F}. Models by \cite{2016A&A...595A..35I} show that low-mass white dwarfs can have rotation rates several times faster than their orbital period upon arrival on the cooling track.

We detected a significant pulsation mode in GD\,278 at $6729.0\pm1.8$\,s, which is so far the longest pulsation period detected in a white dwarf. Previously, SDSS\,J222859.93+362359.6 had the longest detected pulsation period at $6234.9\pm6.0$\,s. That object also happens to be the coolest known pulsating ELM white dwarf with ${T}_{\mathrm{eff}} = 7870 \pm 120$\,K \citep{2013MNRAS.436.3573H}. However, extensive monitoring of SDSS\,J222859.93+362359.6 does not show radial-velocity variability. Unless this object is in an orbit with a low inclination, it may not actually be an ELM white dwarf \citep{2017ApJ...835..180B}. 

We see marginal evidence from a K-S test of all pulsation modes present that there is a roughly 84\,s mean period spacing, which is what we expect from asteroseismic models of a roughly 0.20\,\msun\ ELM white dwarf computed by both MESA and the {\tt LPCODE} stellar evolution codes, assuming the star has a thick hydrogen layer. While we do not adopt a formally best-fit asteroseismic fit with our exploration of a grid of models using {\tt LPCODE}, we do find that there are asteroseismic solutions for our {\em TESS} observations of GD\,278 with parameters that are consistent with those found from spectroscopy \citep{2020ApJ...889...49B}, however further inferred luminosities/distances do not agree. Given the large number of variables that lead to these conclusions, it is difficult to explain the discrepancies.

With continued monitoring of GD\,278, it may be possible to measure rates of change of the significant frequencies within the star, which in general are related to the cooling rate of the white dwarf and the rate of change of the white dwarf radius \citep{2008ARA&A..46..157W}. There are no significant constraints yet on rates of period change for ELM white dwarfs, but these rates of period change would be incredibly valuable at distinguishing whether the ELM white dwarf has undergone CNO flashes in its past \citep{2017A&A...600A..73C}. Unfortunately, GD\,278 will not be observed by \emph{TESS} in Cycle 4, but hopefully it will again in a future extended mission. Starting with Cycle 5, full frame images will be saved every 300\,s, enabling asteroseismic study of even more bright ELM white dwarfs.

\begin{acknowledgments}

We would like to thank Dave Doss, John Kuehne, Coyne Gibson, and the other staff at McDonald Observatory for their support while observing. We would also like to thank the anonymous referee for their helpful feedback.

I.D.L. acknowledges support by NSF grant AST-1812874 and by the Massachusetts Space Grant Consortium (MASGC). We acknowledge salary and travel support through {\em TESS} Guest Investigator Program 80NSSC20K0592, as well as {\em K2} Guest Observer Program 80NSSC19K0162. L.M.C. and A.H.C. acknowledge the AGENCIA through the Programa de Modernizaci\'on Tecnol\'ogica BID 1728/OC-AR, and by the PIP 112-200801-00940 grant from CONICET. K.J.B. is supported by the National Science Foundation under Award AST-1903828.

This work has made use of data from the European Space Agency (ESA) mission {\it Gaia} (\url{https://www.cosmos.esa.int/gaia}), processed by the {\it Gaia} Data Processing and Analysis Consortium (DPAC, \url{https://www.cosmos.esa.int/web/gaia/dpac/consortium}). Funding for the DPAC has been provided by national institutions, in particular the institutions participating in the {\it Gaia} Multilateral Agreement. This paper includes data collected by the \emph{TESS} mission, which are publicly available from  the Mikulski Archive for Space Telescopes. Funding for the \emph{TESS} mission is provided by the NASA Science Mission Directorate. 
\end{acknowledgments}

%

\facilities{\emph{Gaia}, Struve, \emph{TESS}}


\software{IRAF \citep{1986SPIE..627..733T},
        barycorrpy \citep{2018RNAAS...2....4K},
        lightkurve \citep{2018ascl.soft12013L},
        Pyriod \citep{2020AAS...23510606B}, 
        MESA \citep{2019ApJS..243...10P}, 
        GYRE \citep{2013MNRAS.435.3406T},
        matplotlib \citep{2007CSE.....9...90H},
        numpy \citep{2020Natur.585..357H}
        }




\begin{thebibliography}{}
\expandafter\ifx\csname natexlab\endcsname\relax\def\natexlab#1{#1}\fi
\providecommand{\url}[1]{\href{#1}{#1}}
\providecommand{\dodoi}[1]{doi:~\href{http://doi.org/#1}{\nolinkurl{#1}}}
\providecommand{\doeprint}[1]{\href{http://ascl.net/#1}{\nolinkurl{http://ascl.net/#1}}}
\providecommand{\doarXiv}[1]{\href{https://arxiv.org/abs/#1}{\nolinkurl{https://arxiv.org/abs/#1}}}

\bibitem[{{Aerts}(2021)}]{2021RvMP...93a5001A}
{Aerts}, C. 2021, Reviews of Modern Physics, 93, 015001,
  \dodoi{10.1103/RevModPhys.93.015001}

\bibitem[{{Althaus} {et~al.}(2015){Althaus}, {Camisassa}, {Miller Bertolami},
  {C{\'o}rsico}, \& {Garc{\'{\i}}a-Berro}}]{2015A&A...576A...9A}
{Althaus}, L.~G., {Camisassa}, M.~E., {Miller Bertolami}, M.~M., {C{\'o}rsico},
  A.~H., \& {Garc{\'{\i}}a-Berro}, E. 2015, \aap, 576, A9,
  \dodoi{10.1051/0004-6361/201424922}

\bibitem[{{Althaus} {et~al.}(2010){Althaus}, {C{\'o}rsico}, {Isern}, \&
  {Garc{\'\i}a-Berro}}]{2010A&ARv..18..471A}
{Althaus}, L.~G., {C{\'o}rsico}, A.~H., {Isern}, J., \& {Garc{\'\i}a-Berro}, E.
  2010, \aapr, 18, 471, \dodoi{10.1007/s00159-010-0033-1}

\bibitem[{{Althaus} {et~al.}(2013){Althaus}, {Miller Bertolami}, \&
  {C{\'o}rsico}}]{2013A&A...557A..19A}
{Althaus}, L.~G., {Miller Bertolami}, M.~M., \& {C{\'o}rsico}, A.~H. 2013,
  \aap, 557, A19, \dodoi{10.1051/0004-6361/201321868}

\bibitem[{{Althaus} {et~al.}(2001){Althaus}, {Serenelli}, \&
  {Benvenuto}}]{2001MNRAS.323..471A}
{Althaus}, L.~G., {Serenelli}, A.~M., \& {Benvenuto}, O.~G. 2001, \mnras, 323,
  471, \dodoi{10.1046/j.1365-8711.2001.04227.x}

\bibitem[{{Althaus} {et~al.}(2005){Althaus}, {Serenelli}, {Panei},
  {C{\'o}rsico}, {Garc{\'{\i}}a-Berro}, \&
  {Sc{\'o}ccola}}]{2005A&A...435..631A}
{Althaus}, L.~G., {Serenelli}, A.~M., {Panei}, J.~A., {et~al.} 2005, \aap, 435,
  631, \dodoi{10.1051/0004-6361:20041965}

\bibitem[{{Amaro-Seoane} {et~al.}(2017){Amaro-Seoane}, {Audley}, {Babak},
  {Baker}, {Barausse}, {Bender}, {Berti}, {Binetruy}, {Born}, {Bortoluzzi},
  {Camp}, {Caprini}, {Cardoso}, {Colpi}, {Conklin}, {Cornish}, {Cutler},
  {Danzmann}, {Dolesi}, {Ferraioli}, {Ferroni}, {Fitzsimons}, {Gair}, {Gesa
  Bote}, {Giardini}, {Gibert}, {Grimani}, {Halloin}, {Heinzel}, {Hertog},
  {Hewitson}, {Holley-Bockelmann}, {Hollington}, {Hueller}, {Inchauspe},
  {Jetzer}, {Karnesis}, {Killow}, {Klein}, {Klipstein}, {Korsakova}, {Larson},
  {Livas}, {Lloro}, {Man}, {Mance}, {Martino}, {Mateos}, {McKenzie},
  {McWilliams}, {Miller}, {Mueller}, {Nardini}, {Nelemans}, {Nofrarias},
  {Petiteau}, {Pivato}, {Plagnol}, {Porter}, {Reiche}, {Robertson},
  {Robertson}, {Rossi}, {Russano}, {Schutz}, {Sesana}, {Shoemaker}, {Slutsky},
  {Sopuerta}, {Sumner}, {Tamanini}, {Thorpe}, {Troebs}, {Vallisneri},
  {Vecchio}, {Vetrugno}, {Vitale}, {Volonteri}, {Wanner}, {Ward}, {Wass},
  {Weber}, {Ziemer}, \& {Zweifel}}]{2017arXiv170200786A}
{Amaro-Seoane}, P., {Audley}, H., {Babak}, S., {et~al.} 2017, arXiv e-prints,
  arXiv:1702.00786.
\newblock \doarXiv{1702.00786}

\bibitem[{{Bailer-Jones} {et~al.}(2021){Bailer-Jones}, {Rybizki}, {Fouesneau},
  {Demleitner}, \& {Andrae}}]{2021AJ....161..147B}
{Bailer-Jones}, C.~A.~L., {Rybizki}, J., {Fouesneau}, M., {Demleitner}, M., \&
  {Andrae}, R. 2021, \aj, 161, 147, \dodoi{10.3847/1538-3881/abd806}

\bibitem[{{Bell}(2020)}]{2020AAS...23510606B}
{Bell}, K.~J. 2020, in American Astronomical Society Meeting Abstracts, Vol.
  235, American Astronomical Society Meeting Abstracts \#235, 106.06

\bibitem[{{Bell} {et~al.}(2015){Bell}, {Kepler}, {Montgomery}, {Hermes},
  {Harrold}, \& {Winget}}]{2015ASPC..493..217B}
{Bell}, K.~J., {Kepler}, S.~O., {Montgomery}, M.~H., {et~al.} 2015, in
  Astronomical Society of the Pacific Conference Series, Vol. 493, 19th
  European Workshop on White Dwarfs, ed. P.~{Dufour}, P.~{Bergeron}, \&
  G.~{Fontaine}, 217

\bibitem[{{Bell} {et~al.}(2017){Bell}, {Gianninas}, {Hermes}, {Winget},
  {Kilic}, {Montgomery}, {Castanheira}, {Vanderbosch}, {Winget}, \&
  {Brown}}]{2017ApJ...835..180B}
{Bell}, K.~J., {Gianninas}, A., {Hermes}, J.~J., {et~al.} 2017, \apj, 835, 180,
  \dodoi{10.3847/1538-4357/835/2/180}

\bibitem[{{Bell} {et~al.}(2018){Bell}, {Pelisoli}, {Kepler}, {Brown}, {Winget},
  {Winget}, {Vanderbosch}, {Castanheira}, {Hermes}, {Montgomery}, \&
  {Koester}}]{2018A&A...617A...6B}
{Bell}, K.~J., {Pelisoli}, I., {Kepler}, S.~O., {et~al.} 2018, \aap, 617, A6,
  \dodoi{10.1051/0004-6361/201833279}

\bibitem[{{Brown} {et~al.}(2016){Brown}, {Gianninas}, {Kilic}, {Kenyon}, \&
  {Allende Prieto}}]{2016ApJ...818..155B}
{Brown}, W.~R., {Gianninas}, A., {Kilic}, M., {Kenyon}, S.~J., \& {Allende
  Prieto}, C. 2016, \apj, 818, 155, \dodoi{10.3847/0004-637X/818/2/155}

\bibitem[{{Brown} {et~al.}(2013){Brown}, {Kilic}, {Allende Prieto},
  {Gianninas}, \& {Kenyon}}]{2013ApJ...769...66B}
{Brown}, W.~R., {Kilic}, M., {Allende Prieto}, C., {Gianninas}, A., \&
  {Kenyon}, S.~J. 2013, \apj, 769, 66, \dodoi{10.1088/0004-637X/769/1/66}

\bibitem[{{Brown} {et~al.}(2010){Brown}, {Kilic}, {Allende Prieto}, \&
  {Kenyon}}]{2010ApJ...723.1072B}
{Brown}, W.~R., {Kilic}, M., {Allende Prieto}, C., \& {Kenyon}, S.~J. 2010,
  \apj, 723, 1072, \dodoi{10.1088/0004-637X/723/2/1072}

\bibitem[{{Brown} {et~al.}(2012){Brown}, {Kilic}, {Allende Prieto}, \&
  {Kenyon}}]{2012ApJ...744..142B}
---. 2012, \apj, 744, 142, \dodoi{10.1088/0004-637X/744/2/142}

\bibitem[{{Brown} {et~al.}(2020){Brown}, {Kilic}, {Kosakowski}, {Andrews},
  {Heinke}, {Ag{\"u}eros}, {Camilo}, {Gianninas}, {Hermes}, \&
  {Kenyon}}]{2020ApJ...889...49B}
{Brown}, W.~R., {Kilic}, M., {Kosakowski}, A., {et~al.} 2020, \apj, 889, 49,
  \dodoi{10.3847/1538-4357/ab63cd}

\bibitem[{{Calcaferro} {et~al.}(2017{\natexlab{a}}){Calcaferro}, {C{\'o}rsico},
  \& {Althaus}}]{2017A&A...600A..73C}
{Calcaferro}, L.~M., {C{\'o}rsico}, A.~H., \& {Althaus}, L.~G.
  2017{\natexlab{a}}, \aap, 600, A73, \dodoi{10.1051/0004-6361/201630376}

\bibitem[{{Calcaferro} {et~al.}(2018){Calcaferro}, {C{\'o}rsico}, {Althaus},
  {Romero}, \& {Kepler}}]{2018A&A...620A.196C}
{Calcaferro}, L.~M., {C{\'o}rsico}, A.~H., {Althaus}, L.~G., {Romero}, A.~D.,
  \& {Kepler}, S.~O. 2018, \aap, 620, A196, \dodoi{10.1051/0004-6361/201833781}

\bibitem[{{Calcaferro} {et~al.}(2017{\natexlab{b}}){Calcaferro}, {C{\'o}rsico},
  \& {Althaus}}]{2017A&A...607A..33C}
{Calcaferro}, L.~M., {C{\'o}rsico}, A.~H., \& {Althaus}, L. r.~G.
  2017{\natexlab{b}}, \aap, 607, A33, \dodoi{10.1051/0004-6361/201731230}

\bibitem[{{Choi} {et~al.}(2016){Choi}, {Dotter}, {Conroy}, {Cantiello},
  {Paxton}, \& {Johnson}}]{2016ApJ...823..102C}
{Choi}, J., {Dotter}, A., {Conroy}, C., {et~al.} 2016, \apj, 823, 102,
  \dodoi{10.3847/0004-637X/823/2/102}

\bibitem[{{C{\'o}rsico} \& {Althaus}(2006)}]{2006A&A...454..863C}
{C{\'o}rsico}, A.~H., \& {Althaus}, L.~G. 2006, \aap, 454, 863,
  \dodoi{10.1051/0004-6361:20054199}

\bibitem[{{C{\'o}rsico} {et~al.}(2019){C{\'o}rsico}, {Althaus}, {Miller
  Bertolami}, \& {Kepler}}]{2019A&ARv..27....7C}
{C{\'o}rsico}, A.~H., {Althaus}, L.~G., {Miller Bertolami}, M.~M., \& {Kepler},
  S.~O. 2019, \aapr, 27, 7, \dodoi{10.1007/s00159-019-0118-4}

\bibitem[{{Driebe} {et~al.}(1998){Driebe}, {Schoenberner}, {Bloecker}, \&
  {Herwig}}]{1998A&A...339..123D}
{Driebe}, T., {Schoenberner}, D., {Bloecker}, T., \& {Herwig}, F. 1998, \aap,
  339, 123.
\newblock \doarXiv{astro-ph/9809079}

\bibitem[{{Dunlap} {et~al.}(2010){Dunlap}, {Barlow}, \&
  {Clemens}}]{2010ApJ...720L.159D}
{Dunlap}, B.~H., {Barlow}, B.~N., \& {Clemens}, J.~C. 2010, \apjl, 720, L159,
  \dodoi{10.1088/2041-8205/720/2/L159}

\bibitem[{{Fontaine} \& {Brassard}(2008)}]{2008PASP..120.1043F}
{Fontaine}, G., \& {Brassard}, P. 2008, \pasp, 120, 1043,
  \dodoi{10.1086/592788}

\bibitem[{{Fuller} \& {Lai}(2013)}]{2013MNRAS.430..274F}
{Fuller}, J., \& {Lai}, D. 2013, \mnras, 430, 274, \dodoi{10.1093/mnras/sts606}

\bibitem[{{Gaia Collaboration} {et~al.}(2016){Gaia Collaboration}, {Prusti},
  {de Bruijne}, {Brown}, {Vallenari}, {Babusiaux}, {Bailer-Jones}, {Bastian},
  {Biermann}, {Evans}, {Eyer}, {Jansen}, {Jordi}, {Klioner}, {Lammers},
  {Lindegren}, {Luri}, {Mignard}, {Milligan}, {Panem}, {Poinsignon},
  {Pourbaix}, {Randich}, {Sarri}, {Sartoretti}, {Siddiqui}, {Soubiran},
  {Valette}, {van Leeuwen}, {Walton}, {Aerts}, {Arenou}, {Cropper}, {Drimmel},
  {H{\o}g}, {Katz}, {Lattanzi}, {O'Mullane}, {Grebel}, {Holland}, {Huc},
  {Passot}, {Bramante}, {Cacciari}, {Casta{\~n}eda}, {Chaoul}, {Cheek}, {De
  Angeli}, {Fabricius}, {Guerra}, {Hern{\'a}ndez}, {Jean-Antoine-Piccolo},
  {Masana}, {Messineo}, {Mowlavi}, {Nienartowicz}, {Ord{\'o}{\~n}ez-Blanco},
  {Panuzzo}, {Portell}, {Richards}, {Riello}, {Seabroke}, {Tanga},
  {Th{\'e}venin}, {Torra}, {Els}, {Gracia-Abril}, {Comoretto},
  {Garcia-Reinaldos}, {Lock}, {Mercier}, {Altmann}, {Andrae}, {Astraatmadja},
  {Bellas-Velidis}, {Benson}, {Berthier}, {Blomme}, {Busso}, {Carry},
  {Cellino}, {Clementini}, {Cowell}, {Creevey}, {Cuypers}, {Davidson}, {De
  Ridder}, {de Torres}, {Delchambre}, {Dell'Oro}, {Ducourant}, {Fr{\'e}mat},
  {Garc{\'\i}a-Torres}, {Gosset}, {Halbwachs}, {Hambly}, {Harrison}, {Hauser},
  {Hestroffer}, {Hodgkin}, {Huckle}, {Hutton}, {Jasniewicz}, {Jordan},
  {Kontizas}, {Korn}, {Lanzafame}, {Manteiga}, {Moitinho}, {Muinonen},
  {Osinde}, {Pancino}, {Pauwels}, {Petit}, {Recio-Blanco}, {Robin}, {Sarro},
  {Siopis}, {Smith}, {Smith}, {Sozzetti}, {Thuillot}, {van Reeven}, {Viala},
  {Abbas}, {Abreu Aramburu}, {Accart}, {Aguado}, {Allan}, {Allasia},
  {Altavilla}, {{\'A}lvarez}, {Alves}, {Anderson}, {Andrei}, {Anglada Varela},
  {Antiche}, {Antoja}, {Ant{\'o}n}, {Arcay}, {Atzei}, {Ayache}, {Bach},
  {Baker}, {Balaguer-N{\'u}{\~n}ez}, {Barache}, {Barata}, {Barbier}, {Barblan},
  {Baroni}, {Barrado y Navascu{\'e}s}, {Barros}, {Barstow}, {Becciani},
  {Bellazzini}, {Bellei}, {Bello Garc{\'\i}a}, {Belokurov}, {Bendjoya},
  {Berihuete}, {Bianchi}, {Bienaym{\'e}}, {Billebaud}, {Blagorodnova},
  {Blanco-Cuaresma}, {Boch}, {Bombrun}, {Borrachero}, {Bouquillon}, {Bourda},
  {Bouy}, {Bragaglia}, {Breddels}, {Brouillet}, {Br{\"u}semeister},
  {Bucciarelli}, {Budnik}, {Burgess}, {Burgon}, {Burlacu}, {Busonero}, {Buzzi},
  {Caffau}, {Cambras}, {Campbell}, {Cancelliere}, {Cantat-Gaudin}, {Carlucci},
  {Carrasco}, {Castellani}, {Charlot}, {Charnas}, {Charvet}, {Chassat},
  {Chiavassa}, {Clotet}, {Cocozza}, {Collins}, {Collins}, {Costigan}, {Crifo},
  {Cross}, {Crosta}, {Crowley}, {Dafonte}, {Damerdji}, {Dapergolas}, {David},
  {David}, {De Cat}, {de Felice}, {de Laverny}, {De Luise}, {De March}, {de
  Martino}, {de Souza}, {Debosscher}, {del Pozo}, {Delbo}, {Delgado},
  {Delgado}, {di Marco}, {Di Matteo}, {Diakite}, {Distefano}, {Dolding}, {Dos
  Anjos}, {Drazinos}, {Dur{\'a}n}, {Dzigan}, {Ecale}, {Edvardsson}, {Enke},
  {Erdmann}, {Escolar}, {Espina}, {Evans}, {Eynard Bontemps}, {Fabre},
  {Fabrizio}, {Faigler}, {Falc{\~a}o}, {Farr{\`a}s Casas}, {Faye}, {Federici},
  {Fedorets}, {Fern{\'a}ndez-Hern{\'a}ndez}, {Fernique}, {Fienga}, {Figueras},
  {Filippi}, {Findeisen}, {Fonti}, {Fouesneau}, {Fraile}, {Fraser}, {Fuchs},
  {Furnell}, {Gai}, {Galleti}, {Galluccio}, {Garabato}, {Garc{\'\i}a-Sedano},
  {Gar{\'e}}, {Garofalo}, {Garralda}, {Gavras}, {Gerssen}, {Geyer}, {Gilmore},
  {Girona}, {Giuffrida}, {Gomes}, {Gonz{\'a}lez-Marcos},
  {Gonz{\'a}lez-N{\'u}{\~n}ez}, {Gonz{\'a}lez-Vidal}, {Granvik}, {Guerrier},
  {Guillout}, {Guiraud}, {G{\'u}rpide}, {Guti{\'e}rrez-S{\'a}nchez}, {Guy},
  {Haigron}, {Hatzidimitriou}, {Haywood}, {Heiter}, {Helmi}, {Hobbs},
  {Hofmann}, {Holl}, {Holland }, {Hunt}, {Hypki}, {Icardi}, {Irwin}, {Jevardat
  de Fombelle}, {Jofr{\'e}}, {Jonker}, {Jorissen}, {Julbe}, {Karampelas},
  {Kochoska}, {Kohley}, {Kolenberg}, {Kontizas}, {Koposov}, {Kordopatis},
  {Koubsky}, {Kowalczyk}, {Krone-Martins}, {Kudryashova}, {Kull}, {Bachchan},
  {Lacoste-Seris}, {Lanza}, {Lavigne}, {Le Poncin-Lafitte}, {Lebreton},
  {Lebzelter}, {Leccia}, {Leclerc}, {Lecoeur-Taibi}, {Lemaitre}, {Lenhardt},
  {Leroux}, {Liao}, {Licata}, {Lindstr{\o}m}, {Lister}, {Livanou}, {Lobel},
  {L{\"o}ffler}, {L{\'o}pez}, {Lopez-Lozano}, {Lorenz}, {Loureiro},
  {MacDonald}, {Magalh{\~a}es Fernandes}, {Managau}, {Mann}, {Mantelet},
  {Marchal}, {Marchant}, {Marconi}, {Marie}, {Marinoni}, {Marrese},
  {Marschalk{\'o}}, {Marshall}, {Mart{\'\i}n-Fleitas}, {Martino}, {Mary},
  {Matijevi{\v{c}}}, {Mazeh}, {McMillan}, {Messina}, {Mestre}, {Michalik},
  {Millar}, {Miranda}, {Molina}, {Molinaro}, {Molinaro}, {Moln{\'a}r},
  {Moniez}, {Montegriffo}, {Monteiro}, {Mor}, {Mora}, {Morbidelli}, {Morel},
  {Morgenthaler}, {Morley}, {Morris}, {Mulone}, {Muraveva}, {Musella},
  {Narbonne}, {Nelemans}, {Nicastro}, {Noval}, {Ord{\'e}novic},
  {Ordieres-Mer{\'e}}, {Osborne}, {Pagani}, {Pagano}, {Pailler}, {Palacin},
  {Palaversa}, {Parsons}, {Paulsen}, {Pecoraro}, {Pedrosa}, {Pentik{\"a}inen},
  {Pereira}, {Pichon}, {Piersimoni}, {Pineau}, {Plachy}, {Plum}, {Poujoulet},
  {Pr{\v{s}}a}, {Pulone}, {Ragaini}, {Rago}, {Rambaux}, {Ramos-Lerate},
  {Ranalli}, {Rauw}, {Read}, {Regibo}, {Renk}, {Reyl{\'e}}, {Ribeiro},
  {Rimoldini}, {Ripepi}, {Riva}, {Rixon}, {Roelens}, {Romero-G{\'o}mez},
  {Rowell}, {Royer}, {Rudolph}, {Ruiz-Dern}, {Sadowski}, {Sagrist{\`a}
  Sell{\'e}s}, {Sahlmann}, {Salgado}, {Salguero}, {Sarasso}, {Savietto},
  {Schnorhk}, {Schultheis}, {Sciacca}, {Segol}, {Segovia}, {Segransan},
  {Serpell}, {Shih}, {Smareglia}, {Smart}, {Smith}, {Solano}, {Solitro},
  {Sordo}, {Soria Nieto}, {Souchay}, {Spagna}, {Spoto}, {Stampa}, {Steele},
  {Steidelm{\"u}ller}, {Stephenson}, {Stoev}, {Suess}, {S{\"u}veges}, {Surdej},
  {Szabados}, {Szegedi-Elek}, {Tapiador}, {Taris}, {Tauran}, {Taylor},
  {Teixeira}, {Terrett}, {Tingley}, {Trager}, {Turon}, {Ulla}, {Utrilla},
  {Valentini}, {van Elteren}, {Van Hemelryck}, {van Leeuwen}, {Varadi},
  {Vecchiato}, {Veljanoski}, {Via}, {Vicente}, {Vogt}, {Voss}, {Votruba},
  {Voutsinas}, {Walmsley}, {Weiler}, {Weingrill}, {Werner}, {Wevers},
  {Whitehead}, {Wyrzykowski}, {Yoldas}, {{\v{Z}}erjal}, {Zucker}, {Zurbach},
  {Zwitter}, {Alecu}, {Allen}, {Allende Prieto}, {Amorim},
  {Anglada-Escud{\'e}}, {Arsenijevic}, {Azaz}, {Balm}, {Beck}, {Bernstein},
  {Bigot}, {Bijaoui}, {Blasco}, {Bonfigli}, {Bono}, {Boudreault}, {Bressan},
  {Brown}, {Brunet}, {Bunclark}, {Buonanno}, {Butkevich}, {Carret}, {Carrion},
  {Chemin}, {Ch{\'e}reau}, {Corcione}, {Darmigny}, {de Boer}, {de Teodoro}, {de
  Zeeuw}, {Delle Luche}, {Domingues}, {Dubath}, {Fodor}, {Fr{\'e}zouls},
  {Fries}, {Fustes}, {Fyfe}, {Gallardo}, {Gallegos}, {Gardiol}, {Gebran},
  {Gomboc}, {G{\'o}mez}, {Grux}, {Gueguen}, {Heyrovsky}, {Hoar}, {Iannicola},
  {Isasi Parache}, {Janotto}, {Joliet}, {Jonckheere}, {Keil}, {Kim},
  {Klagyivik}, {Klar}, {Knude}, {Kochukhov}, {Kolka}, {Kos}, {Kutka}, {Lainey},
  {LeBouquin}, {Liu}, {Loreggia}, {Makarov}, {Marseille}, {Martayan},
  {Martinez-Rubi}, {Massart}, {Meynadier}, {Mignot}, {Munari}, {Nguyen},
  {Nordlander}, {Ocvirk}, {O'Flaherty}, {Olias Sanz}, {Ortiz}, {Osorio},
  {Oszkiewicz}, {Ouzounis}, {Palmer}, {Park}, {Pasquato}, {Peltzer}, {Peralta},
  {P{\'e}turaud}, {Pieniluoma}, {Pigozzi}, {Poels}, {Prat}, {Prod'homme},
  {Raison}, {Rebordao}, {Risquez}, {Rocca-Volmerange}, {Rosen}, {Ruiz-Fuertes},
  {Russo}, {Sembay}, {Serraller Vizcaino}, {Short}, {Siebert}, {Silva},
  {Sinachopoulos}, {Slezak}, {Soffel}, {Sosnowska}, {Strai{\v{z}}ys}, {ter
  Linden}, {Terrell}, {Theil}, {Tiede}, {Troisi}, {Tsalmantza}, {Tur},
  {Vaccari}, {Vachier}, {Valles}, {Van Hamme}, {Veltz}, {Virtanen}, {Wallut},
  {Wichmann}, {Wilkinson}, {Ziaeepour}, \& {Zschocke}}]{2016A&A...595A...1G}
{Gaia Collaboration}, {Prusti}, T., {de Bruijne}, J.~H.~J., {et~al.} 2016,
  \aap, 595, A1, \dodoi{10.1051/0004-6361/201629272}

\bibitem[{{Gentile Fusillo} {et~al.}(2019){Gentile Fusillo}, {Tremblay},
  {G{\"a}nsicke}, {Manser}, {Cunningham}, {Cukanovaite}, {Hollands}, {Marsh},
  {Raddi}, {Jordan}, {Toonen}, {Geier}, {Barstow}, \&
  {Cummings}}]{2019MNRAS.482.4570G}
{Gentile Fusillo}, N.~P., {Tremblay}, P.-E., {G{\"a}nsicke}, B.~T., {et~al.}
  2019, \mnras, 482, 4570, \dodoi{10.1093/mnras/sty3016}

\bibitem[{{Giammichele} {et~al.}(2017){Giammichele}, {Charpinet}, {Brassard},
  \& {Fontaine}}]{2017A&A...598A.109G}
{Giammichele}, N., {Charpinet}, S., {Brassard}, P., \& {Fontaine}, G. 2017,
  \aap, 598, A109, \dodoi{10.1051/0004-6361/201629935}

\bibitem[{{Gianninas} {et~al.}(2015){Gianninas}, {Kilic}, {Brown}, {Canton}, \&
  {Kenyon}}]{2015ApJ...812..167G}
{Gianninas}, A., {Kilic}, M., {Brown}, W.~R., {Canton}, P., \& {Kenyon}, S.~J.
  2015, \apj, 812, 167, \dodoi{10.1088/0004-637X/812/2/167}

\bibitem[{{Green} {et~al.}(2019){Green}, {Schlafly}, {Zucker}, {Speagle}, \&
  {Finkbeiner}}]{2019ApJ...887...93G}
{Green}, G.~M., {Schlafly}, E., {Zucker}, C., {Speagle}, J.~S., \&
  {Finkbeiner}, D. 2019, \apj, 887, 93, \dodoi{10.3847/1538-4357/ab5362}

\bibitem[{{Guidry} {et~al.}(2021){Guidry}, {Vanderbosch}, {Hermes}, {Barlow},
  {Lopez}, {Boudreaux}, {Corcoran}, {Bell}, {Montgomery}, {Heintz},
  {Castanheira}, {Reding}, {Dunlap}, {Winget}, {Winget}, \&
  {Kuehne}}]{2021ApJ...912..125G}
{Guidry}, J.~A., {Vanderbosch}, Z.~P., {Hermes}, J.~J., {et~al.} 2021, \apj,
  912, 125, \dodoi{10.3847/1538-4357/abee68}

\bibitem[{{Handler} {et~al.}(1997){Handler}, {Pikall}, {O'Donoghue}, {Buckley},
  {Vauclair}, {Chevreton}, {Giovannini}, {Kepler}, {Goode}, {Provencal},
  {Wood}, {Clemens}, {O'Brien}, {Nather}, {Winget}, {Kleinman}, {Kanaan},
  {Watson}, {Nitta}, {Montgomery}, {Klumpe}, {Bradley}, {Sullivan}, {Wu},
  {Marar}, {Seetha}, {Ashoka}, {Mahra}, {Bhat}, {Babu}, {Leibowitz}, {Hemar},
  {Ibbetson}, {Mashal}, {Meistas}, {Dziembowski}, {Pamyatnykh}, {Moskalik},
  {Zola}, {Pajdosz}, {Krzesinski}, {Solheim}, {Bard}, {Massacand}, {Breger},
  {Gelbmann}, {Paunzen}, \& {North}}]{1997MNRAS.286..303H}
{Handler}, G., {Pikall}, H., {O'Donoghue}, D., {et~al.} 1997, \mnras, 286, 303,
  \dodoi{10.1093/mnras/286.2.303}

\bibitem[{{Harris} {et~al.}(2020){Harris}, {Millman}, {van der Walt},
  {Gommers}, {Virtanen}, {Cournapeau}, {Wieser}, {Taylor}, {Berg}, {Smith},
  {Kern}, {Picus}, {Hoyer}, {van Kerkwijk}, {Brett}, {Haldane}, {del R{\'\i}o},
  {Wiebe}, {Peterson}, {G{\'e}rard-Marchant}, {Sheppard}, {Reddy}, {Weckesser},
  {Abbasi}, {Gohlke}, \& {Oliphant}}]{2020Natur.585..357H}
{Harris}, C.~R., {Millman}, K.~J., {van der Walt}, S.~J., {et~al.} 2020, \nat,
  585, 357, \dodoi{10.1038/s41586-020-2649-2}

\bibitem[{{Hermes} {et~al.}(2012){Hermes}, {Montgomery}, {Winget}, {Brown},
  {Kilic}, \& {Kenyon}}]{2012ApJ...750L..28H}
{Hermes}, J.~J., {Montgomery}, M.~H., {Winget}, D.~E., {et~al.} 2012, \apjl,
  750, L28, \dodoi{10.1088/2041-8205/750/2/L28}

\bibitem[{{Hermes} {et~al.}(2013{\natexlab{a}}){Hermes}, {Montgomery},
  {Winget}, {Brown}, {Gianninas}, {Kilic}, {Kenyon}, {Bell}, \&
  {Harrold}}]{2013ApJ...765..102H}
---. 2013{\natexlab{a}}, \apj, 765, 102, \dodoi{10.1088/0004-637X/765/2/102}

\bibitem[{{Hermes} {et~al.}(2013{\natexlab{b}}){Hermes}, {Montgomery},
  {Gianninas}, {Winget}, {Brown}, {Harrold}, {Bell}, {Kenyon}, {Kilic}, \&
  {Castanheira}}]{2013MNRAS.436.3573H}
{Hermes}, J.~J., {Montgomery}, M.~H., {Gianninas}, A., {et~al.}
  2013{\natexlab{b}}, \mnras, 436, 3573, \dodoi{10.1093/mnras/stt1835}

\bibitem[{{Hermes} {et~al.}(2014{\natexlab{a}}){Hermes}, {Brown}, {Kilic},
  {Gianninas}, {Chote}, {Sullivan}, {Winget}, {Bell}, {Falcon}, {Winget},
  {Mason}, {Harrold}, \& {Montgomery}}]{2014ApJ...792...39H}
{Hermes}, J.~J., {Brown}, W.~R., {Kilic}, M., {et~al.} 2014{\natexlab{a}},
  \apj, 792, 39, \dodoi{10.1088/0004-637X/792/1/39}

\bibitem[{{Hermes} {et~al.}(2014{\natexlab{b}}){Hermes}, {G{\"a}nsicke},
  {Koester}, {Bours}, {Townsley}, {Farihi}, {Marsh}, {Littlefair}, {Dhillon},
  {Gianninas}, {Breedt}, \& {Raddi}}]{2014MNRAS.444.1674H}
{Hermes}, J.~J., {G{\"a}nsicke}, B.~T., {Koester}, D., {et~al.}
  2014{\natexlab{b}}, \mnras, 444, 1674, \dodoi{10.1093/mnras/stu1518}

\bibitem[{{Hermes} {et~al.}(2017){Hermes}, {G{\"a}nsicke}, {Kawaler}, {Greiss},
  {Tremblay}, {Gentile Fusillo}, {Raddi}, {Fanale}, {Bell}, {Dennihy}, {Fuchs},
  {Dunlap}, {Clemens}, {Montgomery}, {Winget}, {Chote}, {Marsh}, \&
  {Redfield}}]{2017ApJS..232...23H}
{Hermes}, J.~J., {G{\"a}nsicke}, B.~T., {Kawaler}, S.~D., {et~al.} 2017, \apjs,
  232, 23, \dodoi{10.3847/1538-4365/aa8bb5}

\bibitem[{{Hunter}(2007)}]{2007CSE.....9...90H}
{Hunter}, J.~D. 2007, Computing in Science and Engineering, 9, 90,
  \dodoi{10.1109/MCSE.2007.55}

\bibitem[{{Iben} \& {Livio}(1993)}]{1993PASP..105.1373I}
{Iben}, Icko, J., \& {Livio}, M. 1993, \pasp, 105, 1373, \dodoi{10.1086/133321}

\bibitem[{{Iben} \& {Tutukov}(1986)}]{1986ApJ...311..742I}
{Iben}, Icko, J., \& {Tutukov}, A.~V. 1986, \apj, 311, 742,
  \dodoi{10.1086/164812}

\bibitem[{{Istrate} {et~al.}(2016{\natexlab{a}}){Istrate}, {Fontaine},
  {Gianninas}, {Grassitelli}, {Marchant}, {Tauris}, \&
  {Langer}}]{2016A&A...595L..12I}
{Istrate}, A.~G., {Fontaine}, G., {Gianninas}, A., {et~al.} 2016{\natexlab{a}},
  \aap, 595, L12, \dodoi{10.1051/0004-6361/201629876}

\bibitem[{{Istrate} {et~al.}(2016{\natexlab{b}}){Istrate}, {Marchant},
  {Tauris}, {Langer}, {Stancliffe}, \& {Grassitelli}}]{2016A&A...595A..35I}
{Istrate}, A.~G., {Marchant}, P., {Tauris}, T.~M., {et~al.} 2016{\natexlab{b}},
  \aap, 595, A35, \dodoi{10.1051/0004-6361/201628874}

\bibitem[{{Kanodia} \& {Wright}(2018)}]{2018RNAAS...2....4K}
{Kanodia}, S., \& {Wright}, J. 2018, Research Notes of the American
  Astronomical Society, 2, 4, \dodoi{10.3847/2515-5172/aaa4b7}

\bibitem[{{Kawaler}(1988)}]{1988IAUS..123..329K}
{Kawaler}, S.~D. 1988, in Advances in Helio- and Asteroseismology, ed.
  J.~{Christensen-Dalsgaard} \& S.~{Frandsen}, Vol. 123, 329

\bibitem[{{Kepler} {et~al.}(2007){Kepler}, {Kleinman}, {Nitta}, {Koester},
  {Castanheira}, {Giovannini}, {Costa}, \& {Althaus}}]{2007MNRAS.375.1315K}
{Kepler}, S.~O., {Kleinman}, S.~J., {Nitta}, A., {et~al.} 2007, \mnras, 375,
  1315, \dodoi{10.1111/j.1365-2966.2006.11388.x}

\bibitem[{{Kilic} {et~al.}(2011){Kilic}, {Brown}, {Allende Prieto},
  {Ag{\"u}eros}, {Heinke}, \& {Kenyon}}]{2011ApJ...727....3K}
{Kilic}, M., {Brown}, W.~R., {Allende Prieto}, C., {et~al.} 2011, \apj, 727, 3,
  \dodoi{10.1088/0004-637X/727/1/3}

\bibitem[{{Kilic} {et~al.}(2012){Kilic}, {Brown}, {Allende Prieto}, {Kenyon},
  {Heinke}, {Ag{\"u}eros}, \& {Kleinman}}]{2012ApJ...751..141K}
---. 2012, \apj, 751, 141, \dodoi{10.1088/0004-637X/751/2/141}

\bibitem[{{Kilic} {et~al.}(2015){Kilic}, {Hermes}, {Gianninas}, \&
  {Brown}}]{2015MNRAS.446L..26K}
{Kilic}, M., {Hermes}, J.~J., {Gianninas}, A., \& {Brown}, W.~R. 2015, \mnras,
  446, L26, \dodoi{10.1093/mnrasl/slu152}

\bibitem[{{Kilic} {et~al.}(2007){Kilic}, {Stanek}, \&
  {Pinsonneault}}]{2007ApJ...671..761K}
{Kilic}, M., {Stanek}, K.~Z., \& {Pinsonneault}, M.~H. 2007, \apj, 671, 761,
  \dodoi{10.1086/522228}

\bibitem[{{Kilic} {et~al.}(2018){Kilic}, {Hermes}, {C{\'o}rsico}, {Kosakowski},
  {Brown}, {Antoniadis}, {Calcaferro}, {Gianninas}, {Althaus}, \&
  {Green}}]{2018MNRAS.479.1267K}
{Kilic}, M., {Hermes}, J.~J., {C{\'o}rsico}, A.~H., {et~al.} 2018, \mnras, 479,
  1267, \dodoi{10.1093/mnras/sty1546}

\bibitem[{{Kosakowski} {et~al.}(2020){Kosakowski}, {Kilic}, {Brown}, \&
  {Gianninas}}]{2020ApJ...894...53K}
{Kosakowski}, A., {Kilic}, M., {Brown}, W.~R., \& {Gianninas}, A. 2020, \apj,
  894, 53, \dodoi{10.3847/1538-4357/ab8300}

\bibitem[{{Lightkurve Collaboration} {et~al.}(2018){Lightkurve Collaboration},
  {Cardoso}, {Hedges}, {Gully-Santiago}, {Saunders}, {Cody}, {Barclay}, {Hall},
  {Sagear}, {Turtelboom}, {Zhang}, {Tzanidakis}, {Mighell}, {Coughlin}, {Bell},
  {Berta-Thompson}, {Williams}, {Dotson}, \& {Barentsen}}]{2018ascl.soft12013L}
{Lightkurve Collaboration}, {Cardoso}, J.~V.~d.~M., {Hedges}, C., {et~al.}
  2018, {Lightkurve: Kepler and TESS time series analysis in Python},
  Astrophysics Source Code Library.
\newblock \doeprint{1812.013}

\bibitem[{{Marsh} {et~al.}(1995){Marsh}, {Dhillon}, \&
  {Duck}}]{1995MNRAS.275..828M}
{Marsh}, T.~R., {Dhillon}, V.~S., \& {Duck}, S.~R. 1995, \mnras, 275, 828,
  \dodoi{10.1093/mnras/275.3.828}

\bibitem[{{Montgomery} {et~al.}(2020){Montgomery}, {Hermes}, {Winget},
  {Dunlap}, \& {Bell}}]{2020ApJ...890...11M}
{Montgomery}, M.~H., {Hermes}, J.~J., {Winget}, D.~E., {Dunlap}, B.~H., \&
  {Bell}, K.~J. 2020, \apj, 890, 11, \dodoi{10.3847/1538-4357/ab6a0e}

\bibitem[{{O'Donoghue}(1994)}]{1994MNRAS.270..222O}
{O'Donoghue}, D. 1994, \mnras, 270, 222, \dodoi{10.1093/mnras/270.2.222}

\bibitem[{{Paczynski}(1976)}]{1976IAUS...73...75P}
{Paczynski}, B. 1976, in Structure and Evolution of Close Binary Systems, ed.
  P.~{Eggleton}, S.~{Mitton}, \& J.~{Whelan}, Vol.~73, 75

\bibitem[{{Panei} {et~al.}(2007){Panei}, {Althaus}, {Chen}, \&
  {Han}}]{2007MNRAS.382..779P}
{Panei}, J.~A., {Althaus}, L.~G., {Chen}, X., \& {Han}, Z. 2007, \mnras, 382,
  779, \dodoi{10.1111/j.1365-2966.2007.12400.x}

\bibitem[{{Paxton} {et~al.}(2011){Paxton}, {Bildsten}, {Dotter}, {Herwig},
  {Lesaffre}, \& {Timmes}}]{2011ApJS..192....3P}
{Paxton}, B., {Bildsten}, L., {Dotter}, A., {et~al.} 2011, \apjs, 192, 3,
  \dodoi{10.1088/0067-0049/192/1/3}

\bibitem[{{Paxton} {et~al.}(2013){Paxton}, {Cantiello}, {Arras}, {Bildsten},
  {Brown}, {Dotter}, {Mankovich}, {Montgomery}, {Stello}, {Timmes}, \&
  {Townsend}}]{2013ApJS..208....4P}
{Paxton}, B., {Cantiello}, M., {Arras}, P., {et~al.} 2013, \apjs, 208, 4,
  \dodoi{10.1088/0067-0049/208/1/4}

\bibitem[{{Paxton} {et~al.}(2015){Paxton}, {Marchant}, {Schwab}, {Bauer},
  {Bildsten}, {Cantiello}, {Dessart}, {Farmer}, {Hu}, {Langer}, {Townsend},
  {Townsley}, \& {Timmes}}]{2015ApJS..220...15P}
{Paxton}, B., {Marchant}, P., {Schwab}, J., {et~al.} 2015, \apjs, 220, 15,
  \dodoi{10.1088/0067-0049/220/1/15}

\bibitem[{{Paxton} {et~al.}(2018){Paxton}, {Schwab}, {Bauer}, {Bildsten},
  {Blinnikov}, {Duffell}, {Farmer}, {Goldberg}, {Marchant}, {Sorokina},
  {Thoul}, {Townsend}, \& {Timmes}}]{2018ApJS..234...34P}
{Paxton}, B., {Schwab}, J., {Bauer}, E.~B., {et~al.} 2018, \apjs, 234, 34,
  \dodoi{10.3847/1538-4365/aaa5a8}

\bibitem[{{Paxton} {et~al.}(2019){Paxton}, {Smolec}, {Schwab}, {Gautschy},
  {Bildsten}, {Cantiello}, {Dotter}, {Farmer}, {Goldberg}, {Jermyn}, {Kanbur},
  {Marchant}, {Thoul}, {Townsend}, {Wolf}, {Zhang}, \&
  {Timmes}}]{2019ApJS..243...10P}
{Paxton}, B., {Smolec}, R., {Schwab}, J., {et~al.} 2019, \apjs, 243, 10,
  \dodoi{10.3847/1538-4365/ab2241}

\bibitem[{{Pelisoli} {et~al.}(2018{\natexlab{a}}){Pelisoli}, {Kepler}, \&
  {Koester}}]{2018MNRAS.475.2480P}
{Pelisoli}, I., {Kepler}, S.~O., \& {Koester}, D. 2018{\natexlab{a}}, \mnras,
  475, 2480, \dodoi{10.1093/mnras/sty011}

\bibitem[{{Pelisoli} {et~al.}(2018{\natexlab{b}}){Pelisoli}, {Kepler},
  {Koester}, {Castanheira}, {Romero}, \& {Fraga}}]{2018MNRAS.478..867P}
{Pelisoli}, I., {Kepler}, S.~O., {Koester}, D., {et~al.} 2018{\natexlab{b}},
  \mnras, 478, 867, \dodoi{10.1093/mnras/sty1101}

\bibitem[{{Pelisoli} \& {Vos}(2019)}]{2019MNRAS.488.2892P}
{Pelisoli}, I., \& {Vos}, J. 2019, \mnras, 488, 2892,
  \dodoi{10.1093/mnras/stz1876}

\bibitem[{{Raddi} {et~al.}(2017){Raddi}, {Gentile Fusillo}, {Pala}, {Hermes},
  {G{\"a}nsicke}, {Chote}, {Holland s}, {Henden}, {Catal{\'a}n}, {Geier},
  {Koester}, {Munari}, {Napiwotzki}, \& {Tremblay}}]{2017MNRAS.472.4173R}
{Raddi}, R., {Gentile Fusillo}, N.~P., {Pala}, A.~F., {et~al.} 2017, \mnras,
  472, 4173, \dodoi{10.1093/mnras/stx2243}

\bibitem[{{Ricker} {et~al.}(2015){Ricker}, {Winn}, {Vanderspek}, {Latham},
  {Bakos}, {Bean}, {Berta-Thompson}, {Brown}, {Buchhave}, {Butler}, {Butler},
  {Chaplin}, {Charbonneau}, {Christensen-Dalsgaard}, {Clampin}, {Deming},
  {Doty}, {De Lee}, {Dressing}, {Dunham}, {Endl}, {Fressin}, {Ge}, {Henning},
  {Holman}, {Howard}, {Ida}, {Jenkins}, {Jernigan}, {Johnson}, {Kaltenegger},
  {Kawai}, {Kjeldsen}, {Laughlin}, {Levine}, {Lin}, {Lissauer}, {MacQueen},
  {Marcy}, {McCullough}, {Morton}, {Narita}, {Paegert}, {Palle}, {Pepe},
  {Pepper}, {Quirrenbach}, {Rinehart}, {Sasselov}, {Sato}, {Seager},
  {Sozzetti}, {Stassun}, {Sullivan}, {Szentgyorgyi}, {Torres}, {Udry}, \&
  {Villasenor}}]{2015JATIS...1a4003R}
{Ricker}, G.~R., {Winn}, J.~N., {Vanderspek}, R., {et~al.} 2015, Journal of
  Astronomical Telescopes, Instruments, and Systems, 1, 014003,
  \dodoi{10.1117/1.JATIS.1.1.014003}

\bibitem[{{Shporer} {et~al.}(2010){Shporer}, {Kaplan}, {Steinfadt}, {Bildsten},
  {Howell}, \& {Mazeh}}]{2010ApJ...725L.200S}
{Shporer}, A., {Kaplan}, D.~L., {Steinfadt}, J. D.~R., {et~al.} 2010, \apjl,
  725, L200, \dodoi{10.1088/2041-8205/725/2/L200}

\bibitem[{{Steinfadt} {et~al.}(2010){Steinfadt}, {Bildsten}, \&
  {Arras}}]{2010ApJ...718..441S}
{Steinfadt}, J. D.~R., {Bildsten}, L., \& {Arras}, P. 2010, \apj, 718, 441,
  \dodoi{10.1088/0004-637X/718/1/441}

\bibitem[{{Tassoul}(1980)}]{1980ApJS...43..469T}
{Tassoul}, M. 1980, \apjs, 43, 469, \dodoi{10.1086/190678}

\bibitem[{{Tody}(1986)}]{1986SPIE..627..733T}
{Tody}, D. 1986, in Society of Photo-Optical Instrumentation Engineers (SPIE)
  Conference Series, Vol. 627, \procspie, ed. D.~L. {Crawford}, 733,
  \dodoi{10.1117/12.968154}

\bibitem[{{Townsend} \& {Teitler}(2013)}]{2013MNRAS.435.3406T}
{Townsend}, R.~H.~D., \& {Teitler}, S.~A. 2013, \mnras, 435, 3406,
  \dodoi{10.1093/mnras/stt1533}

\bibitem[{{Tremblay} {et~al.}(2011){Tremblay}, {Bergeron}, \&
  {Gianninas}}]{2011ApJ...730..128T}
{Tremblay}, P.~E., {Bergeron}, P., \& {Gianninas}, A. 2011, \apj, 730, 128,
  \dodoi{10.1088/0004-637X/730/2/128}

\bibitem[{{Tremblay} {et~al.}(2015){Tremblay}, {Gianninas}, {Kilic}, {Ludwig},
  {Steffen}, {Freytag}, \& {Hermes}}]{2015ApJ...809..148T}
{Tremblay}, P.~E., {Gianninas}, A., {Kilic}, M., {et~al.} 2015, \apj, 809, 148,
  \dodoi{10.1088/0004-637X/809/2/148}

\bibitem[{{Winget} \& {Kepler}(2008)}]{2008ARA&A..46..157W}
{Winget}, D.~E., \& {Kepler}, S.~O. 2008, \araa, 46, 157,
  \dodoi{10.1146/annurev.astro.46.060407.145250}

\bibitem[{{Winget} {et~al.}(1991){Winget}, {Nather}, {Clemens}, {Provencal},
  {Kleinman}, {Bradley}, {Wood}, {Claver}, {Frueh}, {Grauer}, {Hine}, {Hansen},
  {Fontaine}, {Achilleos}, {Wickramasinghe}, {Marar}, {Seetha}, {Ashoka},
  {O'Donoghue}, {Warner}, {Kurtz}, {Buckley}, {Brickhill}, {Vauclair}, {Dolez},
  {Chevreton}, {Barstow}, {Solheim}, {Kanaan}, {Kepler}, {Henry}, \&
  {Kawaler}}]{1991ApJ...378..326W}
{Winget}, D.~E., {Nather}, R.~E., {Clemens}, J.~C., {et~al.} 1991, \apj, 378,
  326, \dodoi{10.1086/170434}

\bibitem[{{Wu}(2001)}]{2001MNRAS.323..248W}
{Wu}, Y. 2001, \mnras, 323, 248, \dodoi{10.1046/j.1365-8711.2001.04224.x}

\end{thebibliography}
\end{document}